\documentclass[aps,prx,twocolumn,longbibliography]{revtex4}

\usepackage{amssymb}
\usepackage{amsbsy}
\usepackage{amsmath}
\usepackage{graphicx}
\usepackage{graphics}
\usepackage{setspace}
\usepackage{array}
\usepackage{color}
\usepackage{fontenc}
\usepackage{textcomp}
\usepackage{bm}
\usepackage{float}
\usepackage[bookmarks=false,linkcolor=blue,urlcolor=blue,colorlinks,citecolor=blue]{hyperref}

\newcommand{\vex}[1]{\bm{\mathrm{#1}}}

\newcommand{\T}{\mathsf{T}}

\newcommand{\bsub}{\begin{subequations}}
\newcommand{\esub}{\end{subequations}}

\graphicspath{{../Figures/}}

\begin{document}
\title{Hybrid Dispersion Dirac semimetal and Hybrid Weyl phases in Luttinger Semimetals: A dynamical approach}
\author{Sayed Ali Akbar Ghorashi}
\affiliation{Department of Physics, William $\&$ Mary, Williamsburg, Virginia 23187, USA}

\date{\today}

\newcommand{\be}{\begin{equation}}
\newcommand{\ee}{\end{equation}}
\newcommand{\bea}{\begin{eqnarray}}
\newcommand{\eea}{\end{eqnarray}}
\newcommand{\h}{\hspace{0.30 cm}}
\newcommand{\vs}{\vspace{0.30 cm}}
\newcommand{\n}{\nonumber}
\begin{abstract}
We show that hybrid Dirac and Weyl semimetals can be realized in a three-dimensional Luttinger semimetal with quadratic band touching (QBT). We illustrate this using periodic kicking scheme. In particular, we focus on a momentum-dependent drivings (nonuniform driving) and demonstrate the realization of various hybrid Dirac and Weyl semimetals. We identify a unique \emph{hybrid dispersion Dirac semimetal} with two nodes, where one of the nodes is linear while the other is dispersed quadraticlly. Next, we show that by tilting QBT via periodic driving and in the presence of an external magnetic field, one can realize various single/double \textit{hybrid Weyl semimetals} depending on the strength of external field. Finally, we note that in principle, phases that are found in this work could also be realized by employing the appropriate electronic interactions.
\end{abstract}
\maketitle

\section{Introduction}
Over the last decade topological phases of matter have attracted many attentions for providing tremendous insights both in fundamental and experimental aspects of condensed matter physics \cite{Chiu2016}. In addition to the gapped phases which initially ignited the field, recently discovery of gapless topological phases stimulated many works towards the understanding of nontrivial topology of gapless systems \cite{review1}. Among those, Weyl and Dirac semimetals are of particular interest due to their experimental discovery and many unique physical properties \cite{review1}. The Dirac/Weyl smimetals (DSMs/WSMs) are characterized by isolated point touchings of two degenerate/nondegenerate bands in momentum space. Weyl nodes can be generated by splitting of degenerate Dirac nodes usually via breaking of either time-reversal ($\mathcal{T}$) or inversion ($\mathcal{I}$) symmetries or both. In the standard form, the low energy excitations near the Weyl nodes, disperse linearly along all three momentum directions with each node carries monopole charge of $\pm 1$. As a result, on the surface, there exists a Fermi arc that connects a pair of Weyl nodes with opposite chiralities. Recently, generalization of WSMs to multi-Weyl nodes have been proposed where each nodes have higher-order dispersion in one or more directions and consequently caries monopole charge of larger than one \cite{MultiWeyl1,MultiWeyl2}. \\
\indent On the other hand, DSMs/WSMs can also be classified into type-I and type-II, based on the tilting of their nodes. In the standard type-I WSMs, Fermi surfaces are point-like while in the type-II WSMs, Weyl nodes are tilted resulting in formation of electron and hole pockets producing finite density of states at the Fermi level \cite{typeII1,typeII2}. In a conventional type-I and II WSMs, two Weyl nodes with opposite chiralities have same types, however, recently, a theoretical proposal \cite{hybridWeyloriginal}, introduced a new WSM where a pair of Weyl nodes with different chiralities can have different types, forming the so-called \emph{hybrid Weyl semimetals}.\\
\indent Besides the DSMs/WSMs, another class of three-dimensional nodal semimetals are Luttinger semimetals (LSMs) where possess a quadratic band touching (QBT) point between doubly degenerate valence and conduction bands of $J=3/2$ (effective) fermions at an isolated point in the Brillouin zone.
The LSM provides the low-energy description for a plethora of both strongly and weakly correlated compounds, such as the 227 pyrochlore iridates (Ln2Ir2O7, with Ln being a lanthanide element) \cite{LSMapp1,LSMapp2, LSMapp3,LSMapp4}, half-Heusler compounds (ternary alloys such as LnPtBi, LnPdBi) \cite{LSMapp5,LSMapp6}, HgTe \cite{semiconductor1,semiconductor2,QBToriginal,Ruan2016,mottQBT}, and gray-tin \cite{LSMapp7,LSMapp8}.
Moreover, LSMs proved to show many interesting behaviors\cite{LSMprop1,LSMprop2,LSMprop3,LSMprop4,LSMprop5}, specially in the presence of interaction, for example, investigation of the magnetic and superconducting orders actively have been explored \cite{LSMapp9, Roy-Ghorashi2019, GhorashiPRB2017, GhorashiPRL2018, Ghorashi2019,SatoSC32,LSMBoettcher2,LSMSCLiu1,LSMBoettcher1,Szabo-Bitan32-2018}.\\
\begin{figure*}[ht]
    \centering
    \includegraphics[width=0.6\textwidth]{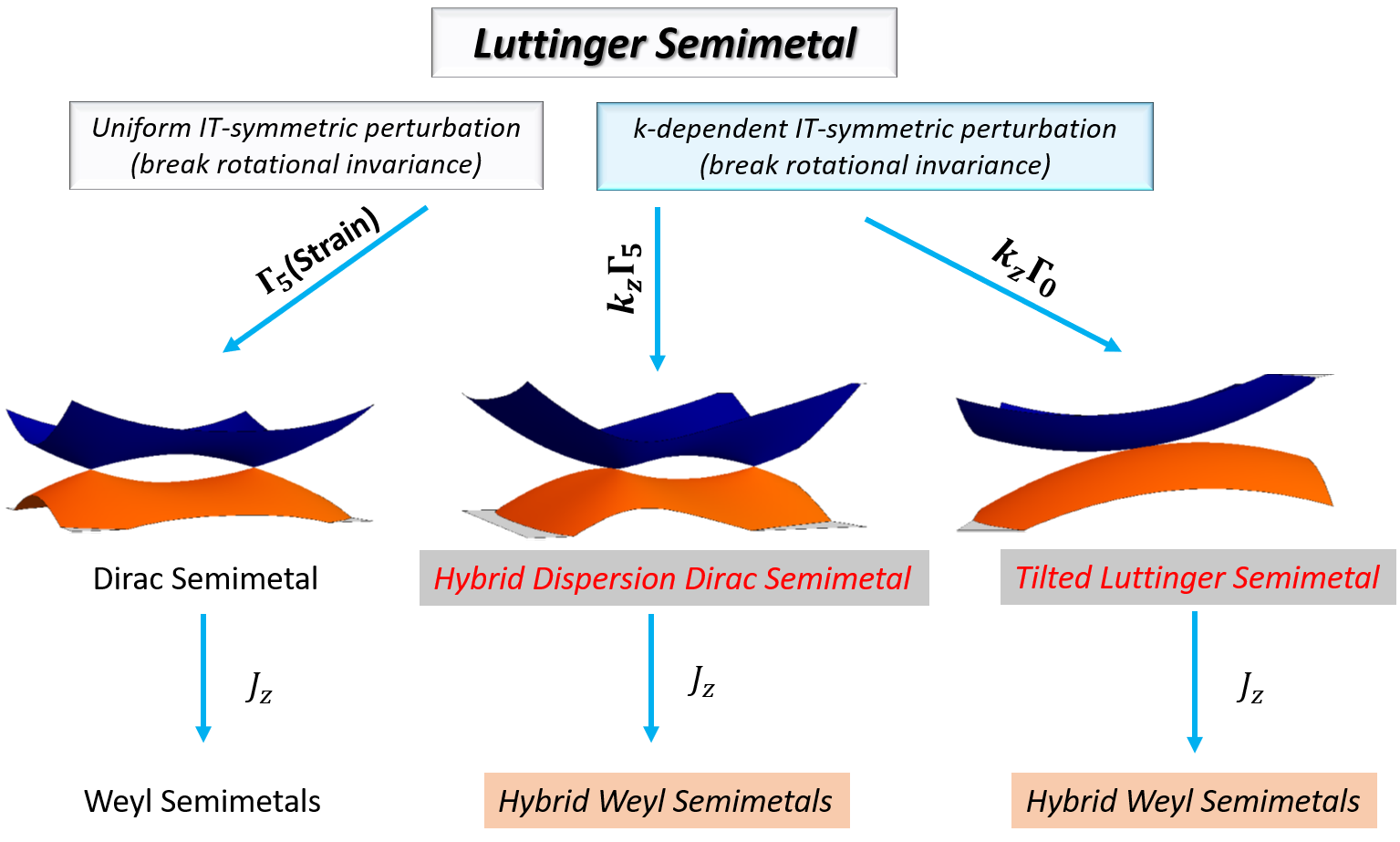}
    \caption{Schematic picture of summary of the results obtained in this work. Starting from a Luttinger semimetal via two different nonuniform ($k$-dependent) periodic kickings, where break inversion ($\mathcal{I}$) and time-reversal ($\mathcal{T}$) symmetries while preserving their combinations ($\mathcal{IT}$), we have obtained a \emph{hybrid dispersion Dirac semimetals} (e.g., $k_z\Gamma_5$ where $\Gamma_i$ are Dirac matrices [see text for details]) and tilted LSM (e.g., $k_z\Gamma_0$). Then, by applying an external magnetic field, $J_z$, in parallel to the direction of nodes (or kick direction), \emph{hybrid Weyl phases} can be generated.}
    \label{fig:adpic}
\end{figure*}
\indent The coexistence of various Weyl nodes with different charges and/or types is an interesting phenomena that could help towards understanding as well as manipulation of the properties of various Weyl nodes in an equal footing setup. Besides a few works reporting the coexistence of type-I and II Dirac/Weyl nodes \cite{hybridWSMexp,hybridDSM1,hybridWSM1,WSMexp1&2,weylcoexPRL}, there have been also proposals\cite{LSMFloq1,Ghorashi2018}, claiming dynamical generation of various Weyl nodes of different types and/or charges in one system. Application of the light is shown to be a powerful method to change the material properties \cite{FloquetRev}. In particular, the conversion of a topologically trivial phase into a nontrivial one using periodic driving has attracted enormous attention in the past decade \cite{FloquetRev,Floquet1, Floquet2, Floquet3}. Specifically many proposals on Floquet WSMs in various systems exist, such as Dirac semimetals \cite{FloqWeyl1,FloqWeyl2}, band insulators \cite{FloqWeyl3}, stacked graphene \cite{FloqWeyl4}, line-nodal semimetals \cite{FloqWeyl2,FloqWeyl5}, and crossing-line semimetals \cite{FloqWeyl6,FloqWeyl7}. Also, proposals have been made to create tunable WSMs in pyrochlore iridates with Zeeman fields \cite{Weylpyro1,Weylpyro2}. Very recently, using circular/elliptic polarized light on Luttinger Hamiltonian in the high-frequency limit we have shown a very rich phase diagram of various Weyl semimetals, including coexistence of type-I and II as well as single and double Weyl nodes \cite{Ghorashi2018} . \\
 \indent Despite the several dynamical proposals for the generation of different Weyl phases, a promising setup for the realization of hybrid Dirac and Weyl semimetals is still lacking. In this work, we tackle this issue by an alternative way of periodic driving, in particular the periodic kicking. Using the periodic $\delta$-function kicks can typically simplify theoretical studies by allowing to perform calculations analytically to a large extent (in contrast to sinusoidal driving or elliptical/circular light) \cite{ Floquet3,kicking2}.
However, for the sake of comparison we also briefly discuss the smooth driving case to show that some of the features of our discussion can be hold up in smooth driving setup as well as long as the perturbation breaks inversion (uniaxially) and time-reversal symmetres but preserve their combinations. In this paper, we show two examples of such perturbations which along with an external magnetic field induce various hybrid Dirac and Weyl phases, including a new \emph{hybrid dispersion Dirac semimetal}. Figure.~\ref{fig:adpic} summarizes the result of this work.

\section{Model and Formalism}
\subsection{Model}
We start with reviewing the main ingredients of Luttinger Hamiltonian in the non-equilibrium limit, which can be represented as,

\begin{align}
    H_{L}(\vec{k})=\int \frac{d^3\vec{k}}{(2\pi)^3}\Psi^{\dagger}_{\vec{k}} h_L(\vec{k}) \Psi_{\vec{k}},
\end{align}
where
\begin{align}
    h_{L}(\vec{k})=&(\frac{k^2}{2m_0}-\mu)\Gamma_0-\frac{1}{2m_1}\sum^3_{a=1} d_a(\vec{k})\Gamma_a\cr
    -&\frac{1}{2m_2}\sum^5_{a=4} d_a(\vec{k})\Gamma_a
\end{align}
where $k^2=k^2_x+k^2_y+k^2_z$ and,
\begin{align}
    \Psi^T_{k}= (c_{\vec{k},3/2}, c_{\vec{k},1/2}, c_{\vec{k},-1/2}, c_{\vec{k},-3/2}).
\end{align}
$\mu$ is the chemical potential measured from the band touching point. $\Gamma_a$ are the well-known gamma matrices which are given by,
\begin{align}
    \Gamma_1=\tau_3\sigma_2,\,\, \Gamma_2=\tau_3\sigma_1,\,\, \Gamma_3=\tau_2,\,\,
    \Gamma_4=\tau_1,\,\, \Gamma_5=\tau_3\sigma_3,
\end{align}
 and satisfy $\{\Gamma_a,\Gamma_b\}=\delta_{a,b}$, while $\Gamma_0$ is four dimensional identity matrix. $\tau$ and $\sigma$ denote space of sign and magnitude of spin projection $m_s\in\{\pm 3/2, \pm 1/2\}$, respectively. $d_a(\vex{k})$ are given as,
 \begin{align}
     d_1=\sqrt{3}k_y k_z,\,\, d_2=\sqrt{3}k_x k_z,\,\, d_3=\sqrt{3}k_x k_y,\,\,\cr
     d_4=\frac{\sqrt{3}}{2}(k_x^2-k_y^2),\,\, d_5=\frac{1}{2}(2k_z^2-k_x^2-k_y^2)
 \end{align}
 The five mutually anticommuting $\Gamma$ matrices can be written
in terms of spin-$3/2$ matrices according to,
\begin{align}
    \Gamma_1=&\frac{1}{\sqrt{3}}\{J_y,J_z\},\,\,\Gamma_2=\frac{1}{\sqrt{3}}\{J_x,J_z\},\,\,\Gamma_3=\frac{1}{\sqrt{3}}\{J_x,J_y\},\cr
    \Gamma_4=&\frac{1}{\sqrt{3}}(J^2_x-J^2_y),\,\,\Gamma_5=J^2_z-\frac{5}{4},
\end{align}
\begin{figure}[t]
\centering
    \includegraphics[width=0.3\textwidth]{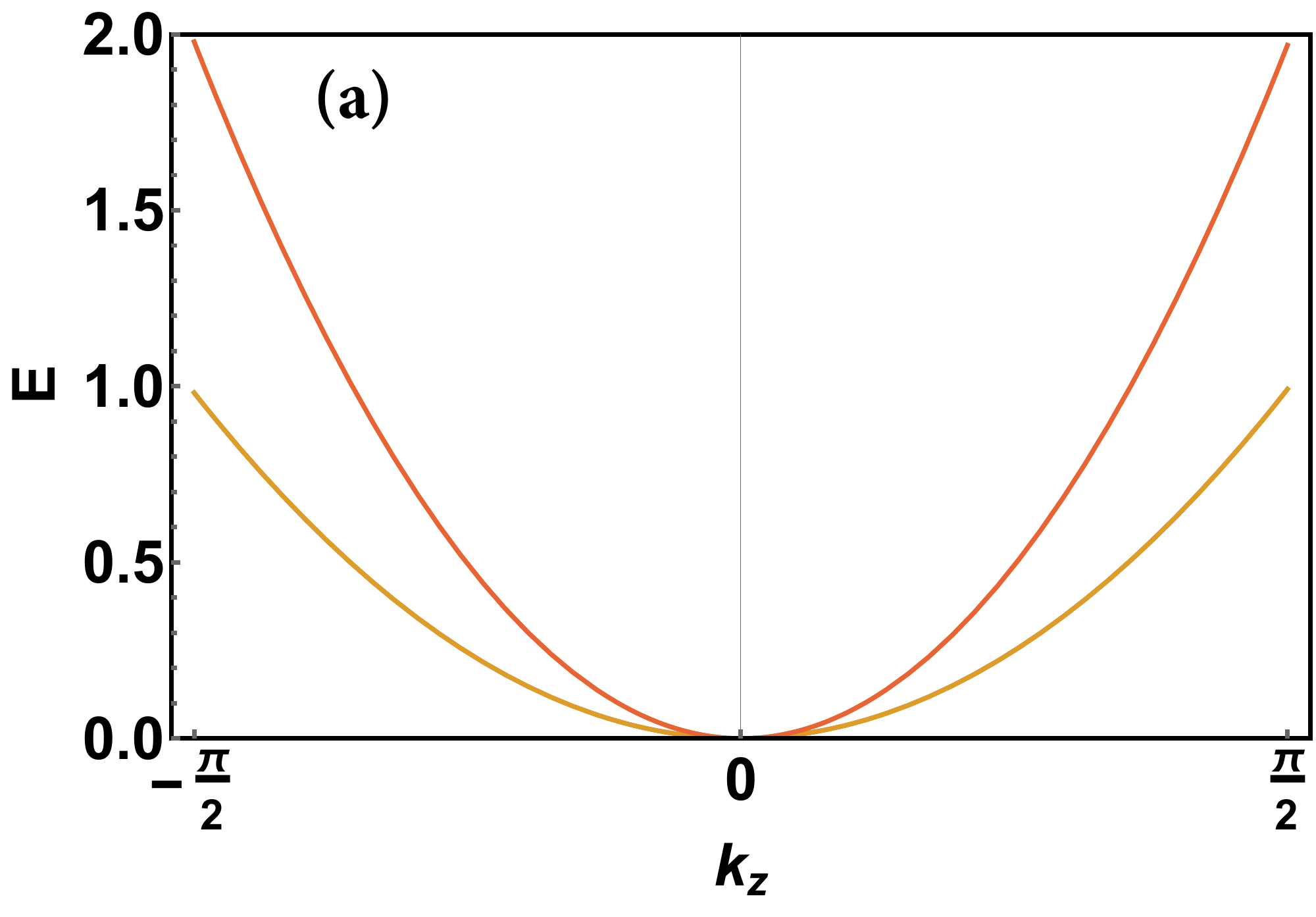}
    \includegraphics[width=0.3\textwidth]{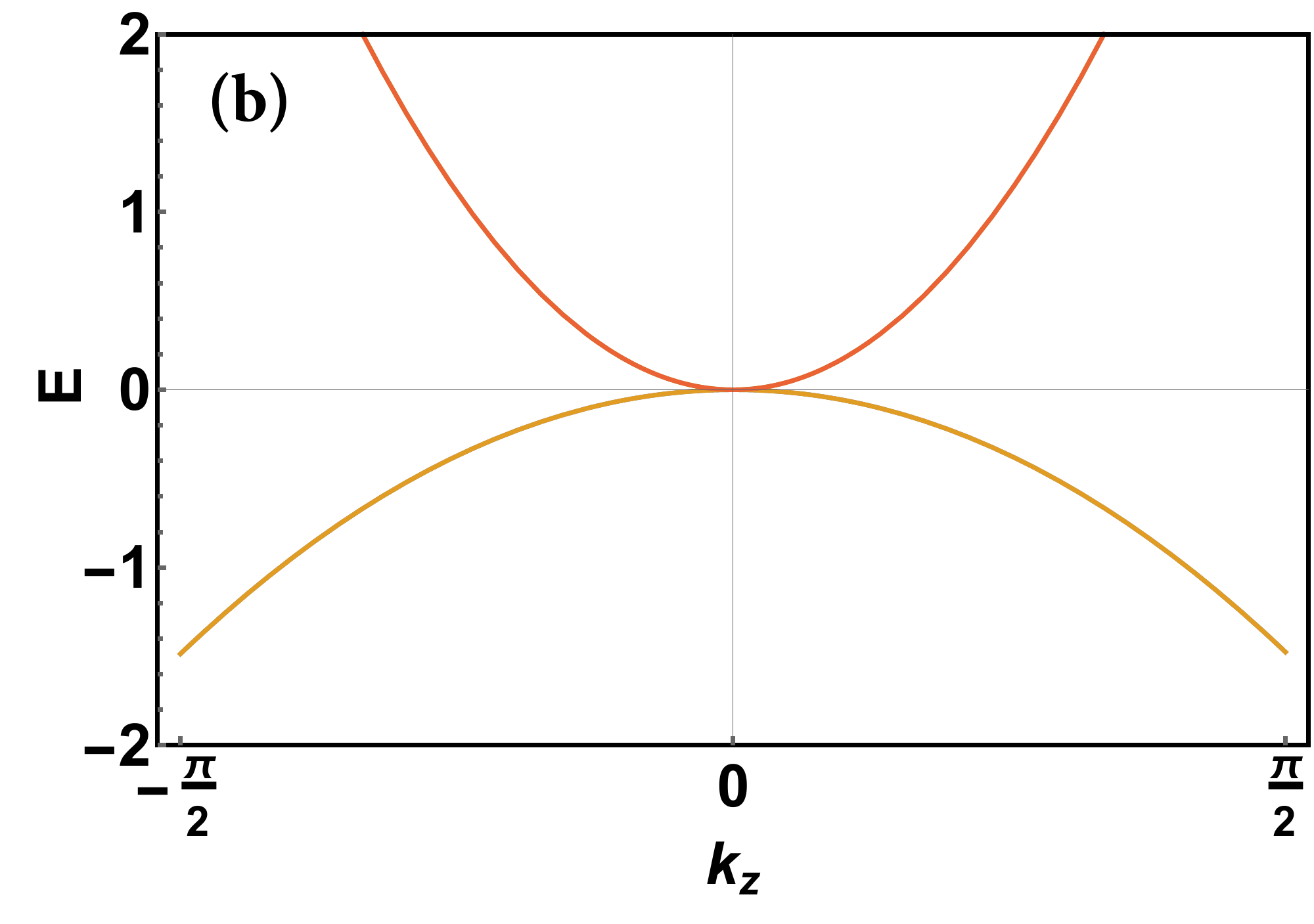}
    \caption{The band structure for LSM along the $k_z$ axis with $\lambda_1=0.6$ and (a) $\lambda_2=0.1$, (b) $\lambda_2=0.6$.}
    \label{fig:LSMbare}
\end{figure}
 Here we take the isotropic limit of $m_1=m_2 \equiv m$. Therefore, the Luttinger Hamiltonian can also be written in an alternative way,
\begin{align}
    h_L(\vec{k})= [(\lambda_1+5\lambda_2/2)k^2-\mu]\Gamma_0-2\lambda_2(\vec{J}.\vec{k})^2
\end{align}
 with $\vec{J}=(J_x,J_y,J_z)$ and $\vec{k}=(k_x,k_y,k_z)$ and we used $\lambda_1=1/2m_0$ and $\lambda_2=1/4m$. $J_{x,y,z}$ are effective spin-$3/2$ operators,
 \begin{eqnarray}
&\,J_z=\begin{bmatrix}
        \frac{3}{2} & 0 & 0 & 0 \\
        0 & \frac{1}{2} & 0 & 0\\
        0 & 0 & -\frac{1}{2} & 0 \\
        0 & 0 & 0 & -\frac{3}{2} \\
      \end{bmatrix},J_x=\begin{bmatrix}
        0 & \frac{\sqrt{3}}{2} & 0 & 0 \\
        \frac{\sqrt{3}}{2} & 0 & 1 & 0\\
        0 & 1 & 0 & \frac{\sqrt{3}}{2} \\
        0 & 0 & \frac{\sqrt{3}}{2} & 0 \\
      \end{bmatrix}\cr
      &\,J_y=\begin{bmatrix}
        0 & \frac{-i\sqrt{3}}{2} & 0 & 0 \\
        \frac{\sqrt{3}}{2} & 0 & -i & 0\\
        0 & 1 & 0 & \frac{-i\sqrt{3}}{2} \\
        0 & 0 & \frac{\sqrt{3}}{2} & 0 \\
      \end{bmatrix}.
\end{eqnarray}
The energy dispersions are $E(k)=(\lambda_{1}\mp2\lambda_{2})k^{2}-\mu$ for the $j=3/2$ and the $j=1/2$ bands, respectively. Four bands come in doubly degenerate pairs as a result of time-reversal (with antiunitary operator $\mathcal{T}=\Gamma_1\Gamma_3\mathcal{K}$ and $\vec{k}\rightarrow -\vec{k}$ where $\mathcal{K}$ is complex conjugation) and inversion ($\mathcal{I}=I_{4\times4}$ and $\vec{k}\rightarrow -\vec{k}$) symmetries. For $\lambda_{2}<2\lambda_{1}$
($\lambda_{2}>2\lambda_{1}$), the degenerate bands curve the same (opposite) way as shown in Figure.~(\ref{fig:LSMbare}). In the case of both bands bending the same way, Eq. (2) is widely used to model heavy- and light-hole bands in zinc-blende semiconductors \cite{semiconductor1} and many properties of such a dispersion have been studied in the literature, including a recent study on the realization of fully gapped topological superconductivity with \emph{p}-wave pairing which has states with exotic cubic and linear dispersions coexisting on the surface \cite{congjunWuPRL16,GhorashiPRB2017}. On the other hand, when bands bend oppositely, the model in Eq. (2) is known as Luttinger semimetal with QBT and is used to describe behavior of certain pyrochlore iridates as well as some doped half-Heusler alloys such as LaPtBi \cite{Chadov2010,Lin2010,halfheusler3}.

\subsection{Periodic driving}
A general time-dependent problem with $H(t)=H_0+V(t)$, can be tackled using Floquet theory when $V(t+T)=V(t)$ is periodic. To proceed, we can expand the periodic potential in a Fourier series as
\begin{align}
    V(t)=V_0+\sum^{\infty}_{n=1} \big(V_n e^{i\omega nt}+V_{-n}e^{-i\omega nt}\big).
\end{align}
 In the limit of fast driving regime, in which the driving frequency is larger than any natural energy scale in the problem, one can obtain the effective Hamiltonian and Floquet operators perturbatively up to  $\mathcal{O}(1/\omega^2)$ \cite{Floquet3}. The Floquet operator, $\mathcal{F}(t)$, is the unitary time-evolution $\hat{U}(t)$ after one period of drive can be factorized as,
\begin{align}
    \mathcal{F}(t)=\exp[-i \alpha\Lambda(\vec{k})]\exp[-i H_L T]=\exp[-i H_{eff} T].
\end{align}
So the dynamics can have three stages: initial kick at $t_i$, the evolution of system with $H(t)$ in the interval $t_f-t_i$ and final kick at $t_f$ which describes the "micromotion" \cite{Floquet3}. Then, the time-evolution operator can be expressed as,
\begin{align}
    \hat{U}(t_i\rightarrow t_f)= \hat{U}(t_f)^{\dagger}e^{-iH_{eff}(t_f-t_i)}\hat{U}(t_i),
\end{align}
where $\hat{U}(t)=e^{-i\mathcal{F}(t)}$. $\mathcal{F}(t)$ is a time-periodic operator with zero average over one period.  Lets set $t_i=0$, then $H_{eff}$ and $\mathcal{F}(t)$ can be expanded as,
\begin{align}
    H_{eff}=\sum_{n=0}^{\infty}\frac{1}{\omega^n}H_{eff}^n,\,\,\mathcal{F}(t)=\sum_{n=1}^{\infty}\frac{1}{\omega^n}\mathcal{F}^n.
\end{align}
A generic quantum $\delta$-kick can be implemented with following perturbation,
\begin{align}
    H_{kick}(t)=\alpha \Lambda(\vec{k}) \sum_{n=-\infty}^{\infty} \delta(t-nT)
\end{align}
\begin{figure}
\centering
    \includegraphics[width=0.3\textwidth]{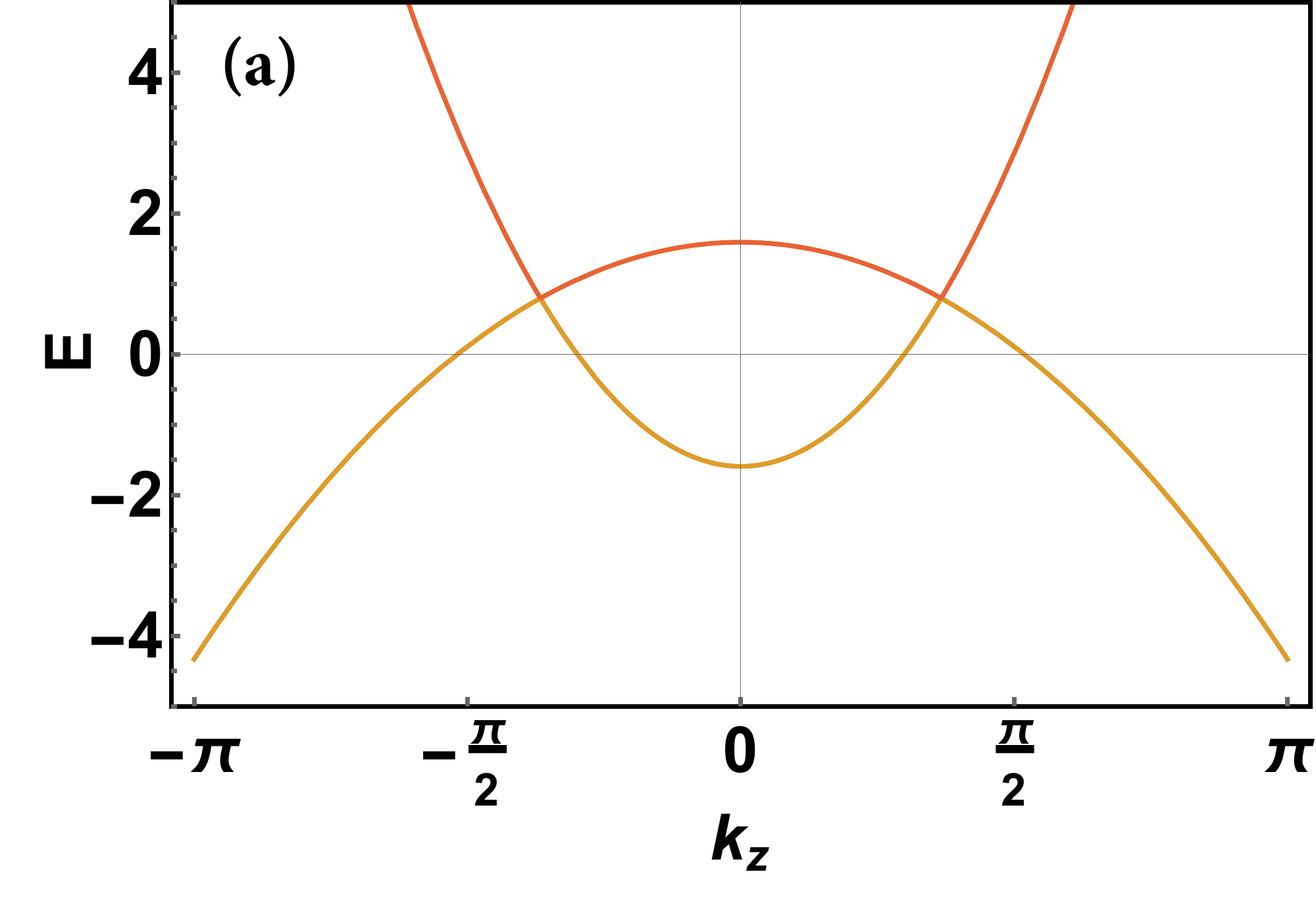}
    \includegraphics[width=0.3\textwidth]{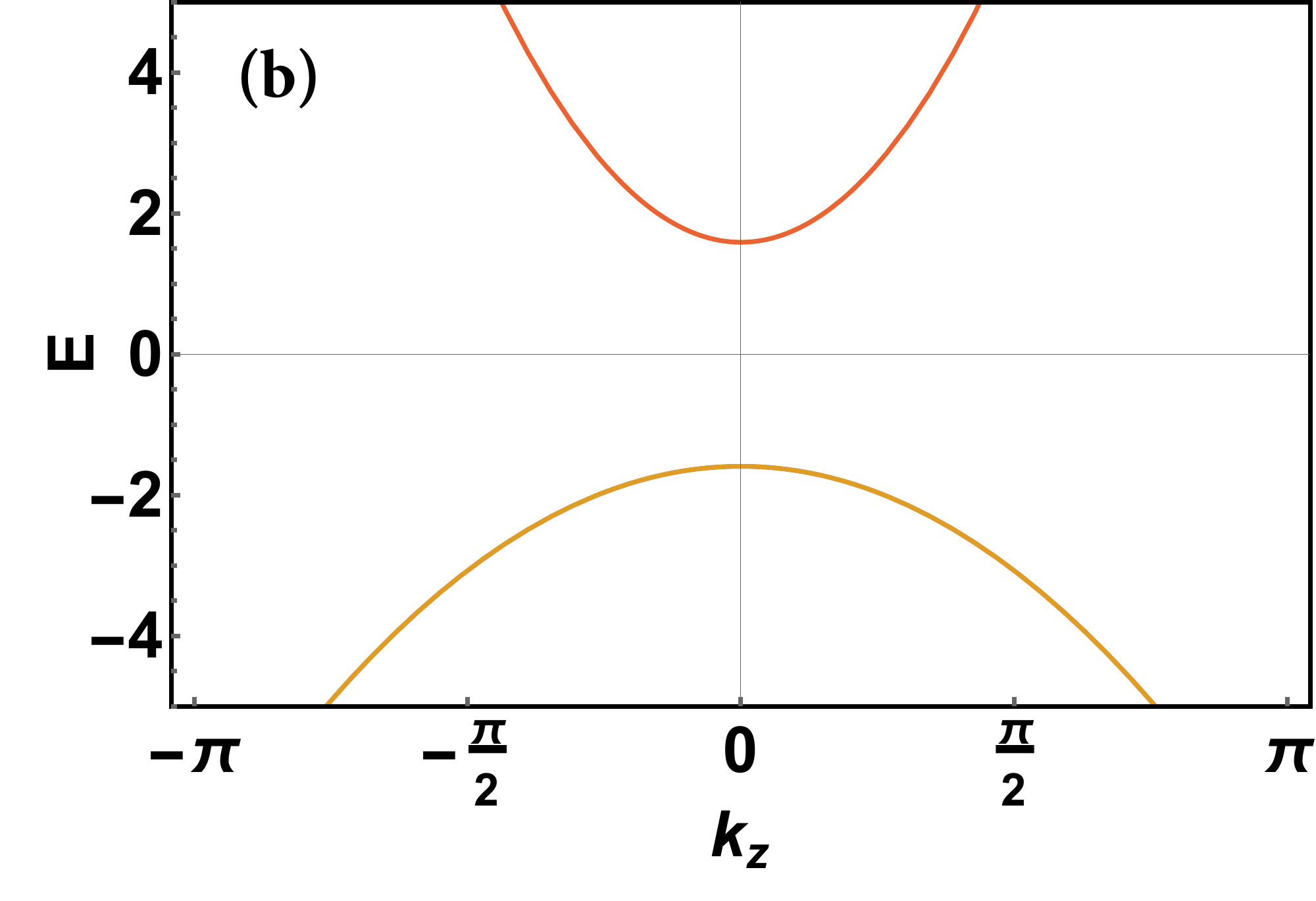}
    \caption{The band structure of LSM along the $k_z$ axis in the presence of uniform strain ($\alpha \Gamma_5$) with (a) $\alpha=0.5$ (tensile) and (b) $\alpha=-0.5$ (compression). $\lambda_1=\lambda_2=0.6$ and $\omega=20$ are used.}
    \label{fig:G5}
\end{figure}
where $T=2\pi/\omega$, $\alpha$ is the kicking strength and $\Lambda(\vec{k})$ is the matrix representation of a perturbation, which in general can be a function of momentum, $\vec{k}$, and could be used to mimic a nonuniform kicking. Following a perturbative expansion the effective Hamiltonian for $\delta$-kick, with $H_{kick,n}=\alpha \Lambda(\vec{k})/T\,\text{for all}\,\,n$, can be obtained as \cite{Floquet3,kicking2},
\begin{align}\label{effkick}
    H^{kick}_{eff}=&\,H_L+\frac{\alpha\Lambda(\vec{k})}{T}+\frac{1}{24}[[\alpha\Lambda(\vec{k}),H_L],\alpha\Lambda(\vec{k})]+\mathcal{O}(1/\omega^3).
\end{align}
We note that the third term in Eq.~(\ref{effkick}), can be dropped in the limit of weak kicks, i.e, to the first order of kick strength. Therefore, a dynamical kicking effectively is equivalent to directly adding the kick term as a perturbation to the bare Hamiltonian, something which is not always relevant in the equilibrium. In the rest of this work we only focus on the physics of the effective Hamiltonian and we leave "micromotion" physics for future studies. \\
Similarly, the effective Hamiltonian for the case of smooth driving of the form $H(t)=H_0+\alpha \Lambda(\vex{k})\cos(\omega t)$ can be obtained as \cite{Floquet3},
\begin{align}\label{effsmooth}
    H^{cos}_{eff}=H_0+\frac{1}{4\omega^2}[[\alpha \Lambda(\vec{k}),H_0],\alpha \Lambda(\vec{k})]+\mathcal{O}(1/\omega^3).
\end{align}
Unlike Eq.~(\ref{effkick}) of periodic kicking, the first correction to the $H_0$ in the case of smooth driving is $\mathcal{O}(\alpha^2)$.\\
\indent In the following we first investigate the effect of uniform kicking ($k$-independent) and show that some of the known results can be retrieved. Then we turn to nonuniform periodic kicking and show that by introducing $k$-dependent $\mathcal{IT}$ symmetric kicks, interestingly,  different hybrid Dirac/Weyl semimetals can be obtained. Finally, we briefly compare the case of smooth driving of Eq.~(\ref{effsmooth}) with the results obtained by periodic kicking.

 \section{Uniform kicking}
There are many possibilities of uniform kicking, including: five $\Gamma_j$, and ten commutators $\Gamma_{ij}=[\Gamma_i,\Gamma_j]/(2i)$ with $i > j$ , which can be expressed in terms of the products of odd number of spin-$3/2$ matrices. Here we focus on uniform kicks which are proportional to $\Lambda=\Gamma_j$. The effective Hamiltonian for such kicks, in a sufficiently weak kick limit, is given by,
\begin{align}\label{effG5}
    H^j_{eff}=h_L(\vec{k})+\frac{\alpha}{T}\Gamma_j,
\end{align}
with spectrum,
\begin{align}
    E^j_{\pm}(\vec{k})=\lambda_1k^2\pm \frac{\sqrt{4\lambda_2^2\sum_i d^2_i T^2 - 4\lambda_2\alpha T d_j+\alpha^2}}{T},
\end{align}

Of particular interest is the $\Gamma_5$ kicking, that breaks the rotational invariance and can be thought of as the effect of an external strain in $z$ direction \cite{Ruan2016,QBToriginal}. It is known that a uniaxial strain, can realize two different phases in LSMs depending on the sign of $\alpha$ \cite{Ruan2016,QBToriginal,Weylpyro1,Weylpyro2,LSMmagnetic}. Figure.~\ref{fig:G5}a(b) shows the bandstructure corresponding to effective Hamiltonian of Eq.~(\ref{effG5}) with $\Gamma_j=\Gamma_5$ and for $\alpha > 0$ ($\alpha<0$) which represents a Dirac semimetal (trivial/topological insulator).

In the case of a tensile strain, there are two Dirac nodes at $k_z=\pm\frac{\alpha}{2\lambda_2 T}$ ($\alpha > 0 $). Now if we add an external magnetic field, we expect that each Dirac nodes split to two Weyl points and form Weyl semimetal phases due to breaking of $\mathcal{T}$ symmetry. Interestingly, even in the topological insulator limit of $\alpha<0$, one can realize Weyl semimetals depending on the strength of magnetic field \cite{LSMmagnetic}. We also note that magnetic field alone can drive a Luttinger semimetal to a Weyl semimetal phase \cite{LSMmagnetic,Weylpyro2}. Moreover, in the experimentally relevant system of pyrochlore irridates, it is shown that magnetic field can give rise to rich phase diagram of topological semimetals \cite{Weylpyro2}. It is important to mention that an external magnetic field can have orbital effects for sufficiently large strength. However, here and in the rest of this paper, we neglect the orbital effects of magnetic field for sake of simplicity following many works in the literature \cite{Weylpyro1, Weylpyro2,LSMFloq1} and we refer readers to some of the works which explored the effects of magnetic field in Landau levels and quantum oscillations \cite{LSMmagnetic,LSMquantumoscillation}. The detailed analysis of the effect of magnetic field (also possible pseudomagnetic fields due to other perturbations) will be left for future studies.

\section{Nonuniform kicking}
Now lets turn to the main part of this work, where we investigate the effect of nonuniform kicks. Similar to the uniform kicking discussed in the previous section, we can consider numerous possibilities of nonuniform driving. However, here, we restrict ourselves to a "linear in momentum" kicks of the form $k_i\Gamma_i$ where $\Gamma_i$ are time-reversal even. The very first consequence of such choice of the kicks, is the breaking of both $\mathcal{T}$ and $\mathcal{I}$ symmetries, while preserving their combinations $\mathcal{IT}$, which is a necessary symmetry for keeping double degeneracy of the bands intact. Specifically, we focus on two types of nonuniform kicking: (a) $k_z\Gamma_5$ and (b) a $k_z\Gamma_0$.

\begin{figure*}[htb]
 \centering
    \includegraphics[width=0.42\textwidth]{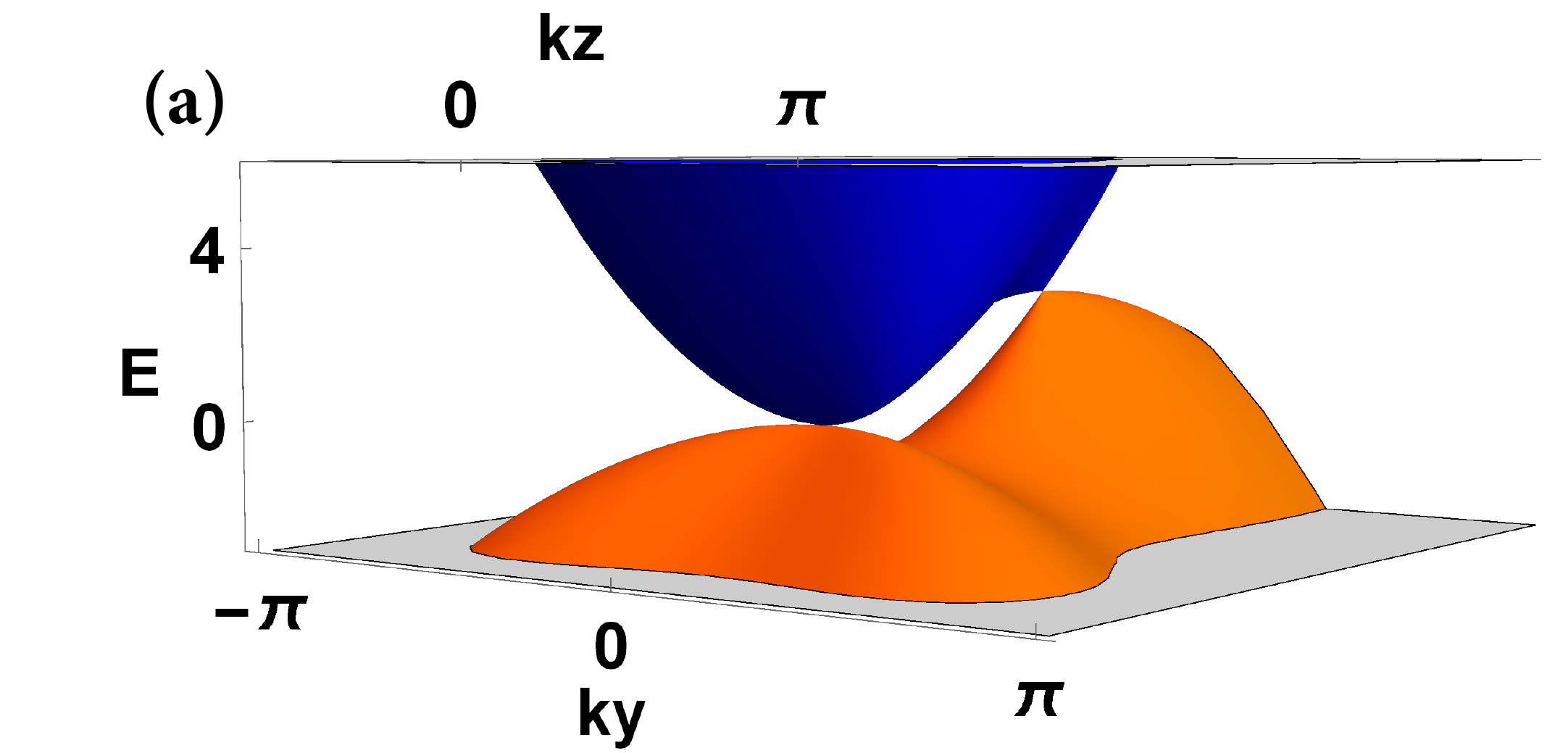}\includegraphics[width=0.3\textwidth]{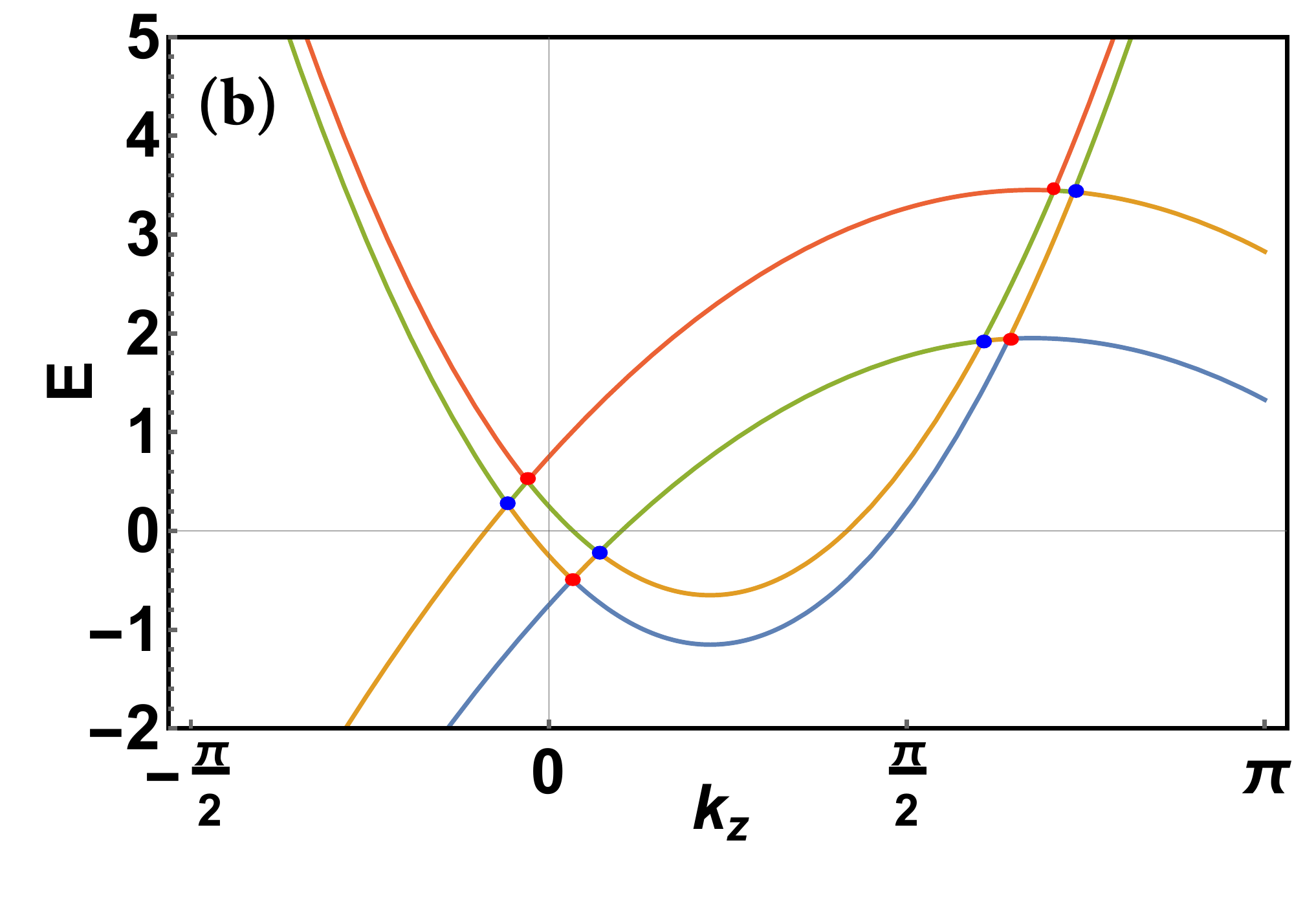}\includegraphics[width=0.3\textwidth]{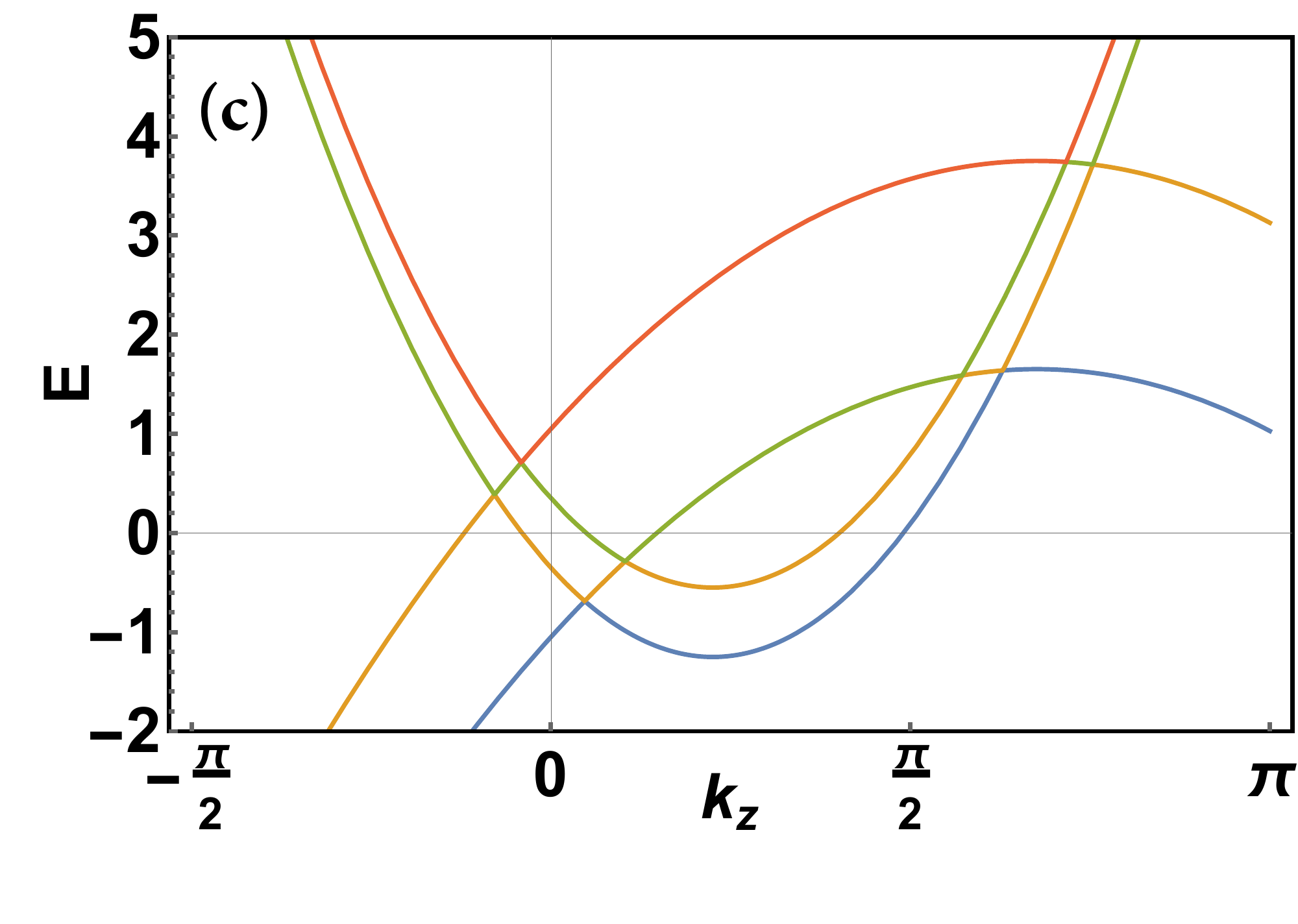}
    \includegraphics[width=0.3\textwidth]{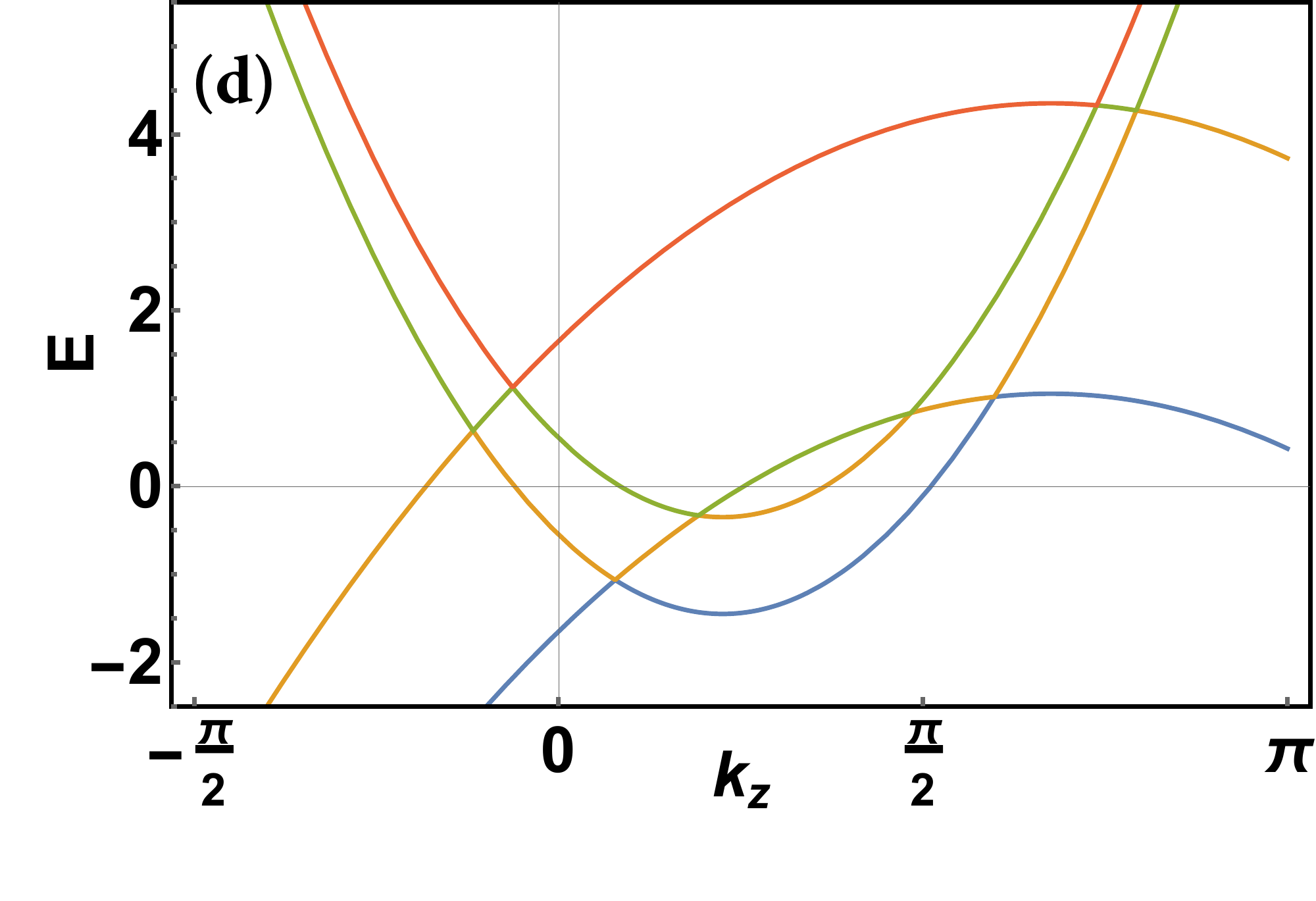}\includegraphics[width=0.3\textwidth]{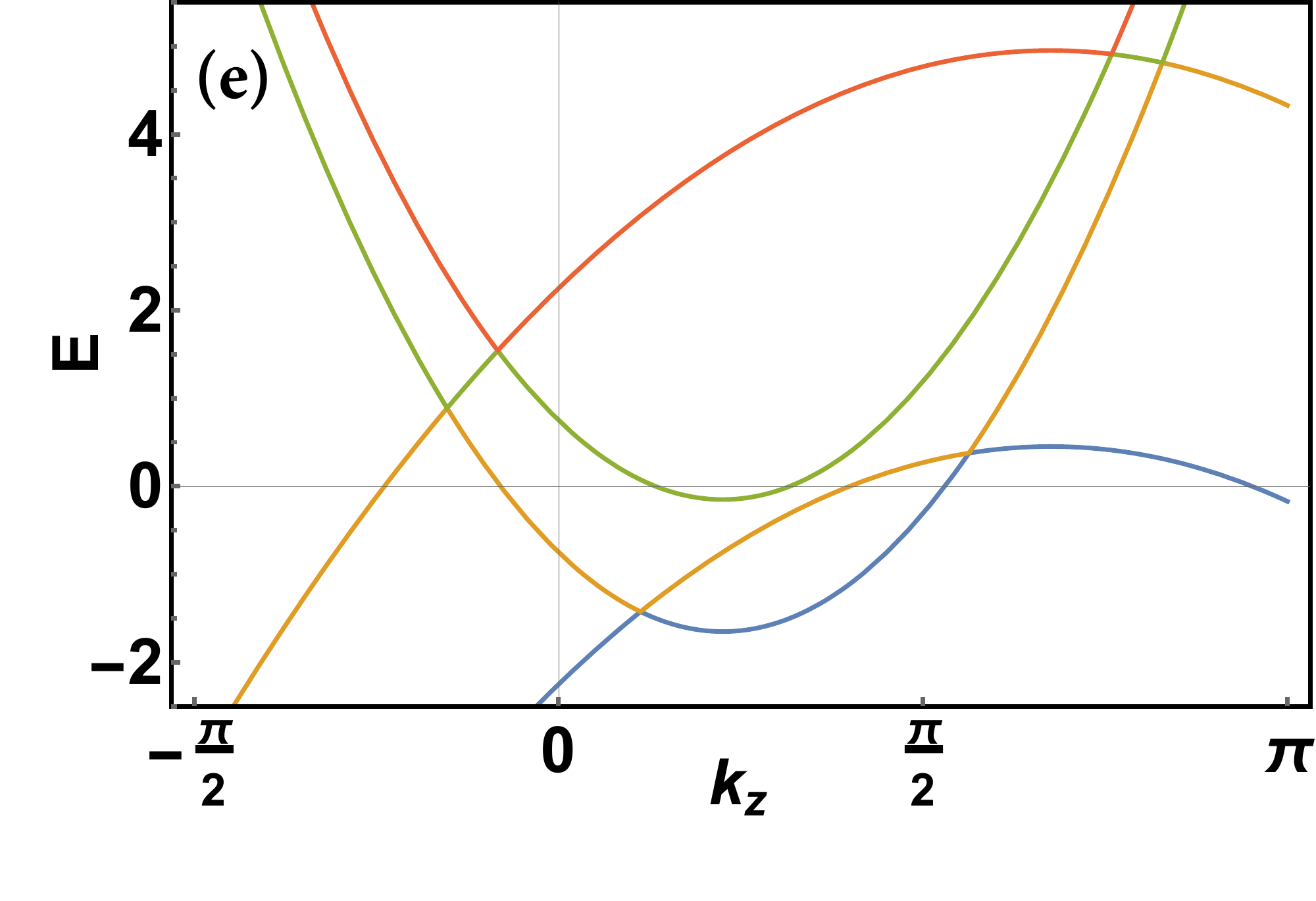}\includegraphics[width=0.3\textwidth]{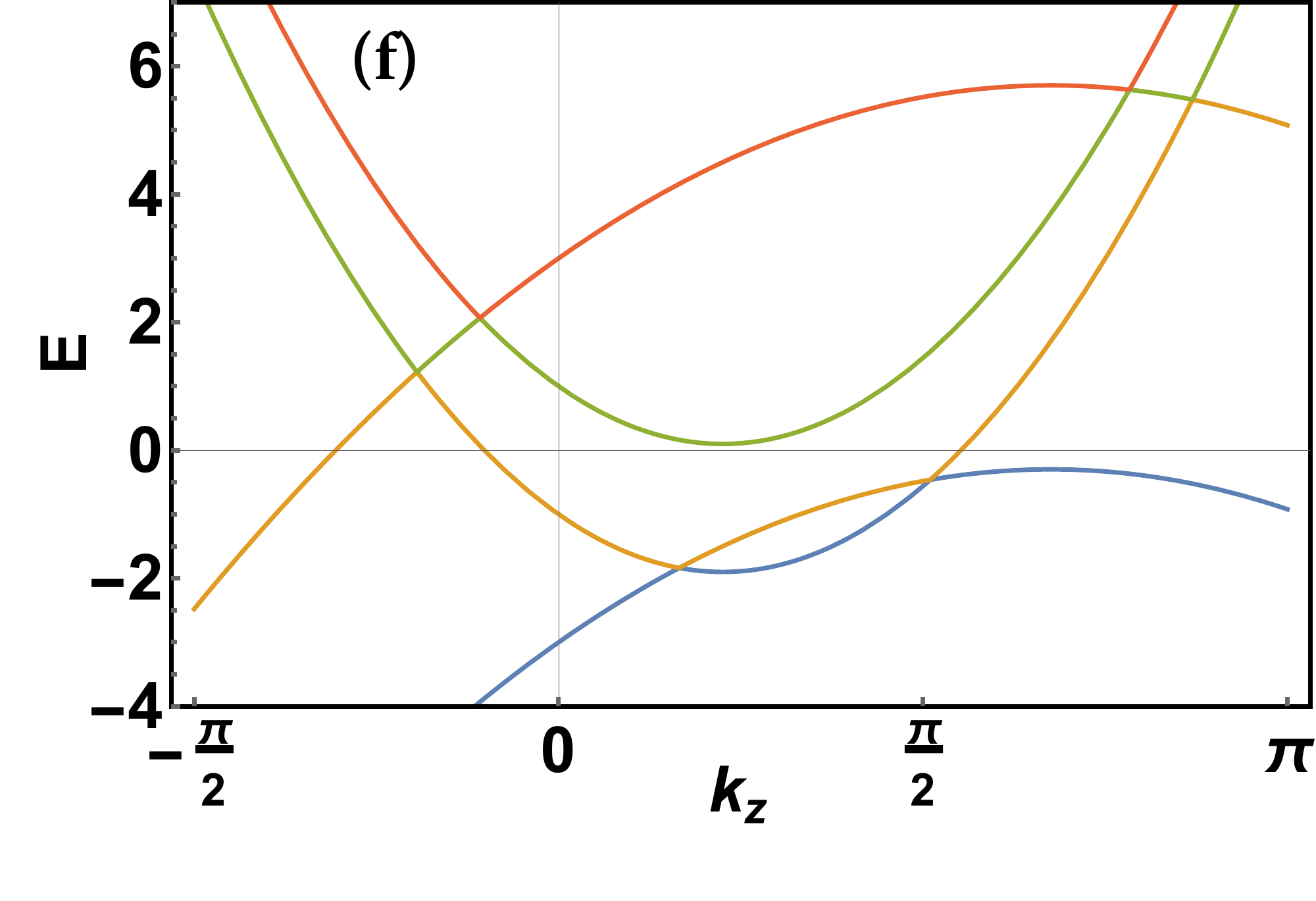}
    \includegraphics[width=0.3\textwidth]{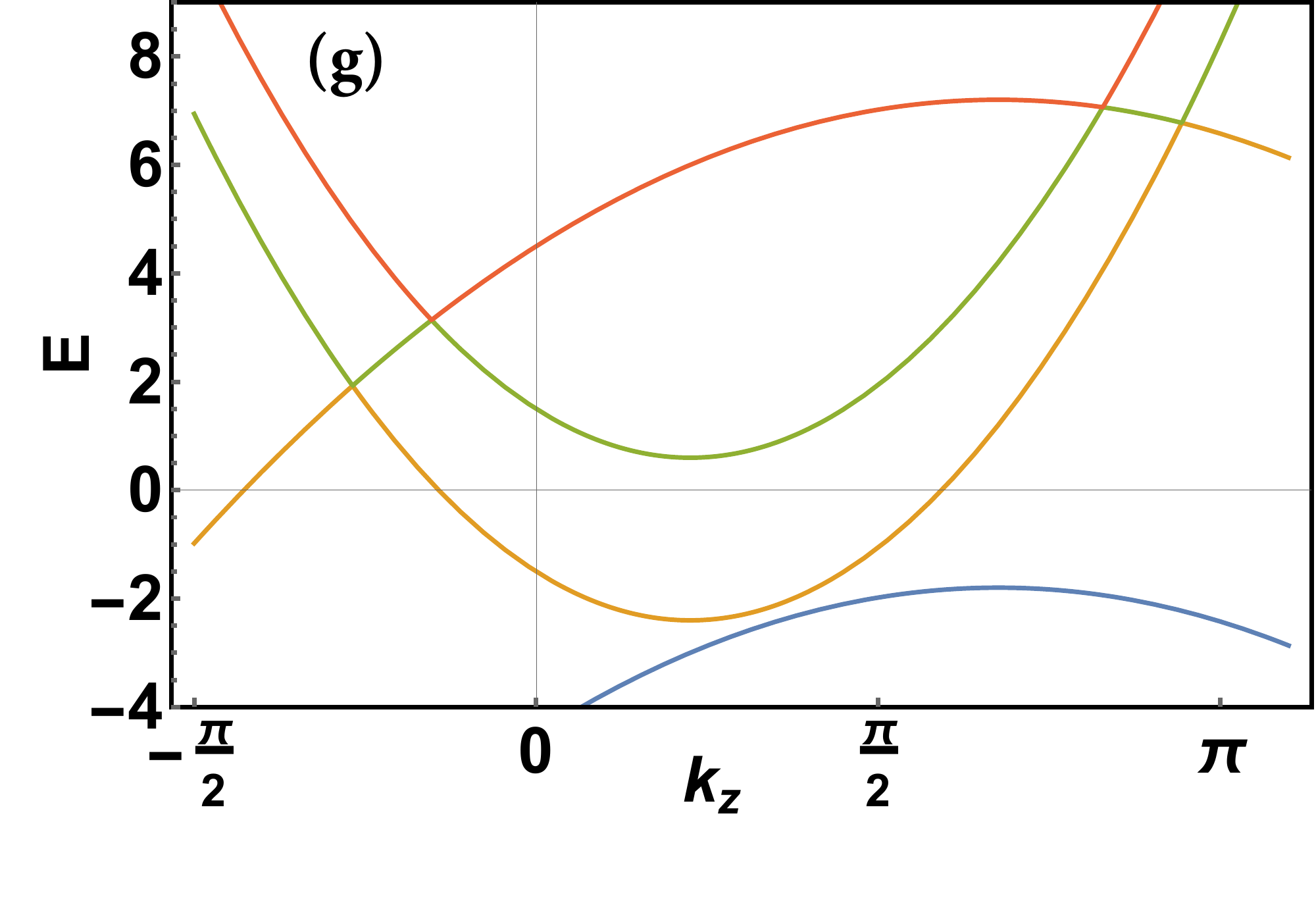}\includegraphics[width=0.42\textwidth]{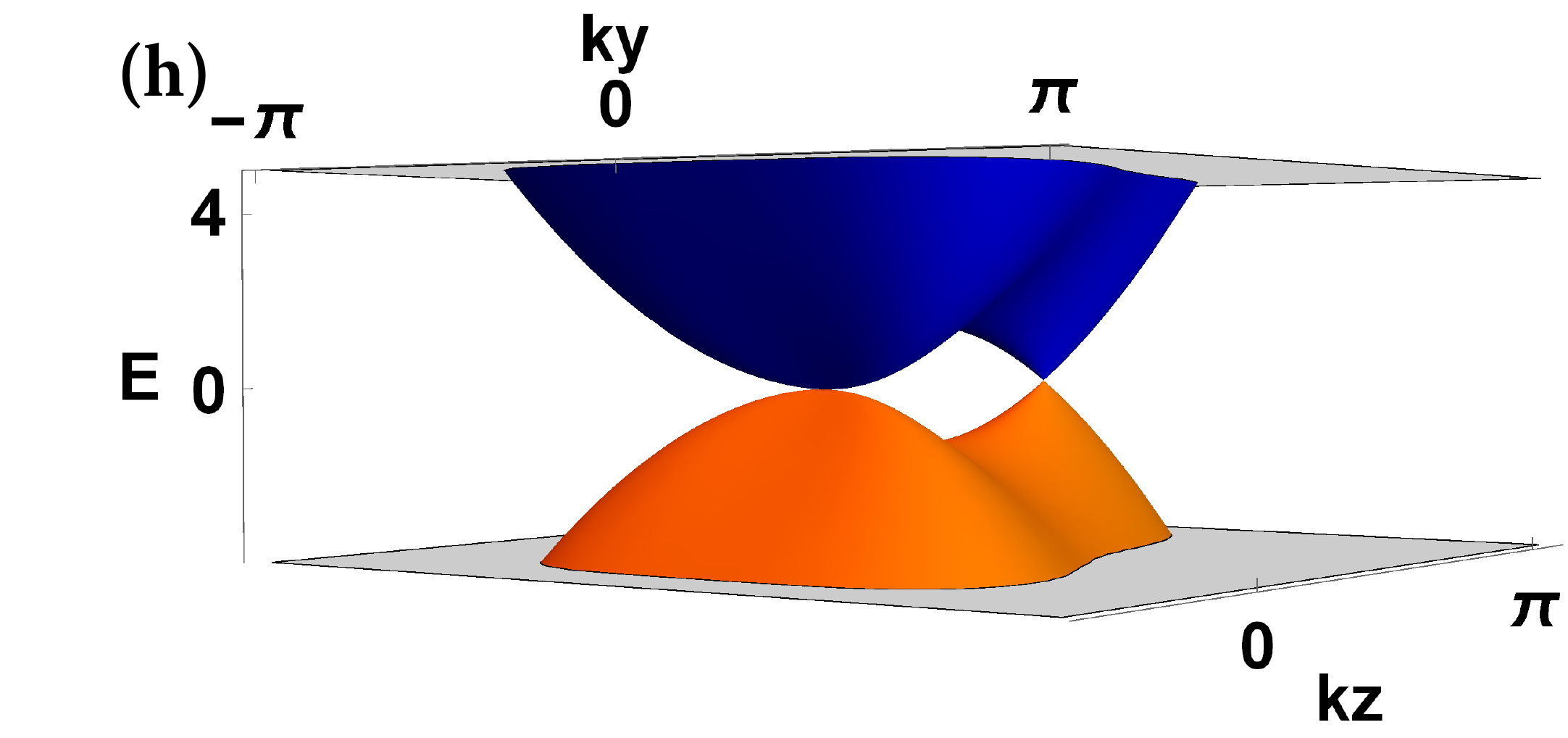}
    \caption{The band structure of LSM in the presence of $k_z\Gamma_5$ with kick strength of $\alpha=0.8$  and (a) $h=0$, $k_y-k_z$ plane showing quadratic dispersion along the $k_y$ for the node at $\Gamma$ point, (b) $h=0.5$, (c) $h=0.7$, (d) $h=1.1$, (e) $h=1.5$, (f) $h=2$, (g) $h=3$ and (h) same as (a) but with $\lambda_1=0$. $\lambda_1=\lambda_2=0.6$ and $\omega=20$ are used in all plots (except (h) where $\lambda_1=0$).}
    \label{fig:kzG5}
\end{figure*}
\subsection{$\Lambda=k_z\Gamma_5$}

As it is mentioned above the $\Gamma_5$ can be interpreted as the effect of strain in $k_z$ directions \cite{Ruan2016}, similarly, $k_z\Gamma_5$ also breaks the rotational invariance, then in principle we might still think of $k_z\Gamma_5$ term as a type of nonuniform strain or strain gradient, even though full microscopic derivation of such strain could shed more light. The presence of $k_z$ suggests that microscopically on a lattice, $\Gamma_5$ now acts non-local and directly modifies hopping terms. On the hand, one can convert the $k_z\Gamma_5$ to the lattice model and explicitly write it down as, $\sum_{\sigma\in\{1/2,3/2\}}(C^{\dagger}_{k_{\perp},i,\sigma,\uparrow}C_{k_{\perp},j,\sigma,\uparrow}-C^{\dagger}_{k_{\perp},i,\sigma,\downarrow}C_{k_{\perp},j,\sigma,\downarrow})/2i+h.c$. This is similar to the spin current operator along $z$-axis. Therefore, we can also think of the $k_z\Gamma_5$ kick as an applied spin current (or related to it) along $z$ direction. However, it is essential to note that besides the rotational invariance the $k_z\Gamma_5$ kick also breaks inversion and time-reversal symmetries while preserving their combinations. Then the bands of driven system are still degenerate. Therefore, here, without restricting ourselves to any particular microscopic interpretation, we emphasize that to achieve the results obtained in this section one need perturbations which break both $\mathcal{I}$ and $\mathcal{T}$ but preserve $\mathcal{IT}$.\\
The effective Hamiltonian for $k_z\Gamma_5$ kick can be written as, $H_{eff}=h_{L}(\vec{k})+\frac{\alpha}{T}k_z\Gamma_5$, with following spectrum,
\begin{align}
    E_{\pm}(\vec{k})=\lambda_1k^2\pm \frac{\sqrt{4\lambda_2^2T^2\sum_i d^2_i - 4\lambda_2\alpha T k_z d_5+\alpha^2 k^2_z}}{T}.
\end{align}
Figure.~\ref{fig:kzG5}, shows the spectrum in $k_x-k_z$ plane. There are two Dirac nodes at $k_z=0$ and $k_z=\frac{\alpha}{2\lambda_2 T}$. While one of the nodes is always pinned to $\Gamma$ point the other node can be tuned by kick parameters. Therefore, for fixed set of parameters the distance between two nodes for $k_z\Gamma_5$ is half of the case with uniform strain $\Gamma_5$. Moreover, we observe three main differences in compare to the case with uniform strain as is shown in Figure.~\ref{fig:G5}. (i) There is no major difference between the tensile ($\alpha >0 $) and  compressive ($\alpha <0 $) strain. In both cases system drives into a Dirac semimetal phase. (ii) Unlike the uniform tensile strain, due to broken inversion symmetry two Dirac nodes reside at different energies and (iii) the most important of all, remarkably, one of the nodes is quadratic while the other is linear, realizing an unique \emph{hybrid Dirac semimetal}. Interestingly, the "quadratic" Dirac node (or a QBT which is linearly dispersed in $k_z$ direction) is pinned at the $\Gamma$ point and the "energy difference" as well as distance between the nodes can be controlled with kick strength. Moreover, while the linear Dirac node is tilted, the node centered around the $\Gamma$ point no matter how strong is the kick, shows no tilt. It should be emphasized, here, we define hybrid Dirac semimetal mainly based on the different dispersion of two Dirac nodes instead of their tilts, even though as we discussed above, they show different tilting (and types) as well. Moreover, We note that such hybrid phase is unique to the Dirac semimetals because it is impossible to realize a Weyl semimetal with pair of nodes with different dispersion or (magnitude of monopole charges).\\
\indent Next, lets apply an external magnetic field $h J_z$ where $h$ denotes the strength of magnetic field. The magnetic field splits Dirac nodes to Weyl nodes in $k_z$ direction. In the presence of a magnetic field an analytical expression of eigenvalues for the general $\lambda_1,\lambda_2$ can not be achieved, so in the following we proceed by solving the model numerically. Figure.~\ref{fig:kzG5} depicts the evolution of the Weyl nodes by strength of external magnetic field. Starting from $h \ll \alpha $, there are 8 nodes on $k_z$ axis, four single and four double Weyl nodes, which are indicated in Figure.~\ref{fig:kzG5} by the red and blue dots, respectively. Interestingly, we observe that one or more of the Weyl pairs realize a \emph{hybrid Weyl phases} and they can survive up to a decently strong magnetic field (Figure.~\ref{fig:kzG5}(b-f)). While the other Weyl nodes do not possess hybrid types, they show different tilts for each nodes of the same pair. Only at a very strong magnetic field (Figure.~\ref{fig:kzG5}(g)) all of the Weyl points show a type-I structure.
Moreover, we see that by increasing the magnetic field, from initial eight nodes, four of them merge and gap out and we left off with only four nodes (two single and two double nodes) at $h>>\alpha$ limit. \\
\indent Finally, we comment on the particle-hole symmetric limit of $\lambda_1=0$. In this limit we can get energies analytically even in presence of an external magnetic field as,
\begin{align}
    E^{(1)}_{\pm}(\vec{k})=&\pm\frac{h}{2}+(2\lambda_2 k^2_z-\frac{\alpha k_z}{T}),\cr E^{(2)}_{\pm}(\vec{k})=&\pm\frac{3h}{2}-(2\lambda_2 k^2_z-\frac{\alpha k_z}{T}),
\end{align}
with 8 nodes located at,
\begin{align}
    k^{\pm,\pm}_{z,I}=&\,\frac{\alpha\pm \sqrt{\alpha^2\pm 4h\lambda_2T^2}}{4\lambda_2 T},\cr
    k^{\pm,\pm}_{z,II}=&\,\frac{\alpha\pm \sqrt{\alpha^2\pm 8h\lambda_2T^2}}{4\lambda_2 T}.
\end{align}
As is clear from the above equations, four out of the eight nodes, $k^{\pm,-}_{z,I,II}$, gap out by increasing $h$. However, it is noteworthy that even though a \emph{hybrid dispersion Dirac semimetal phase} can still be generated, but now in the limit of $\lambda_1=0$, the two nodes have the same energies Figure.~\ref{fig:kzG5}(h), therefore, no longer a hybrid Weyl phase can be achieved in the presence of an external magnetic field.

\begin{figure*}[htb]
\centering
\includegraphics[width=0.3\textwidth]{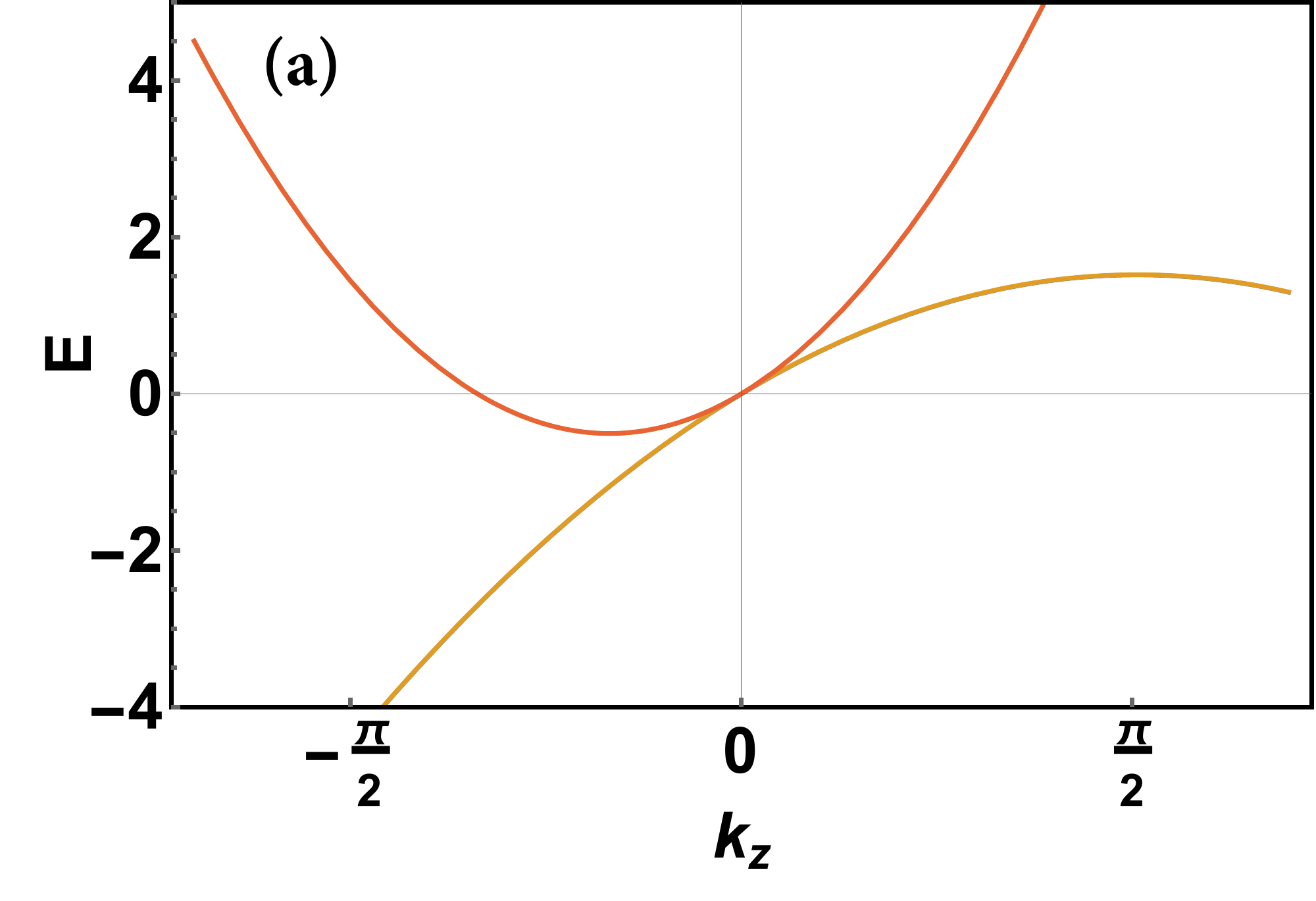}\includegraphics[width=0.3\textwidth]{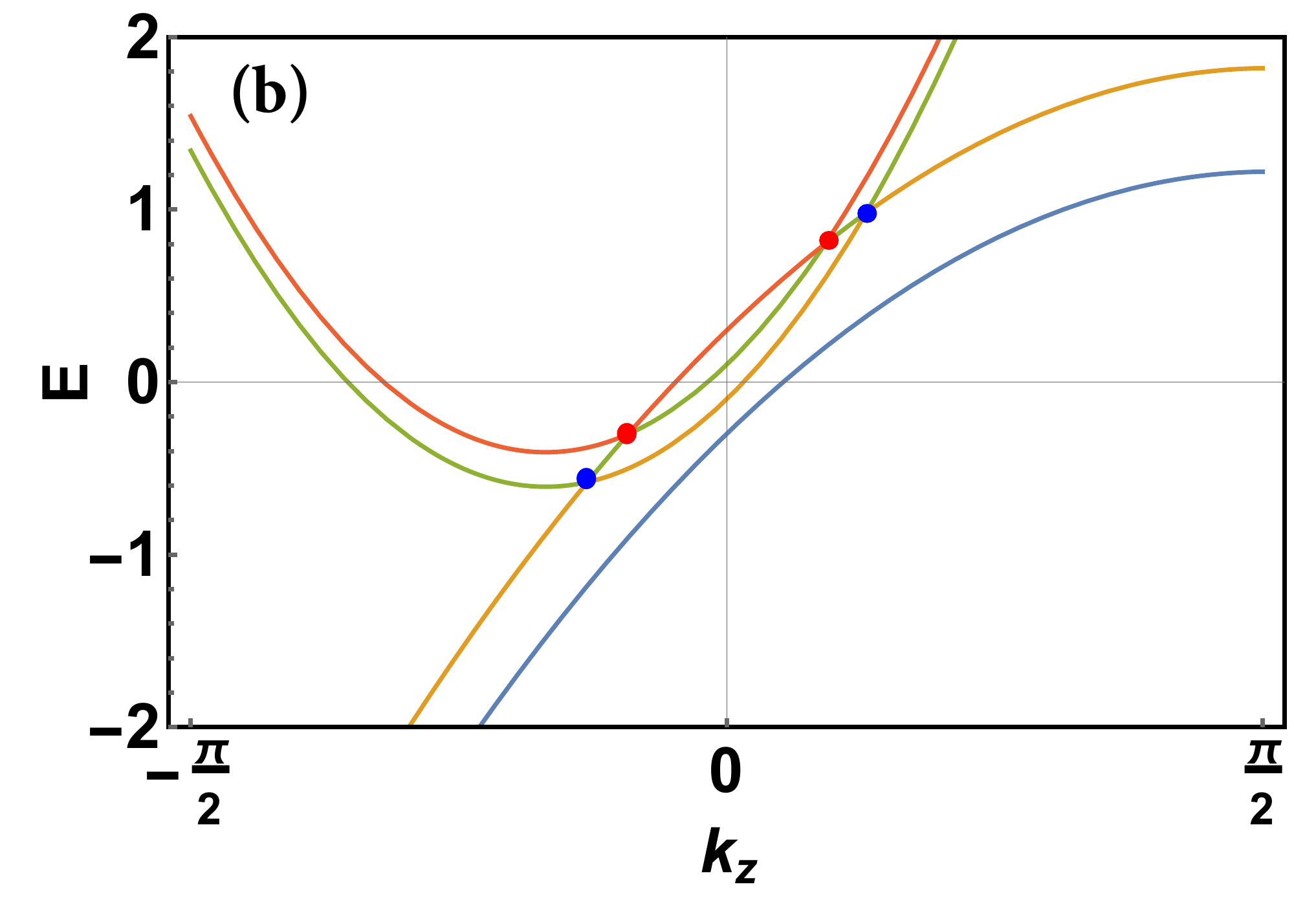}\includegraphics[width=0.3\textwidth]{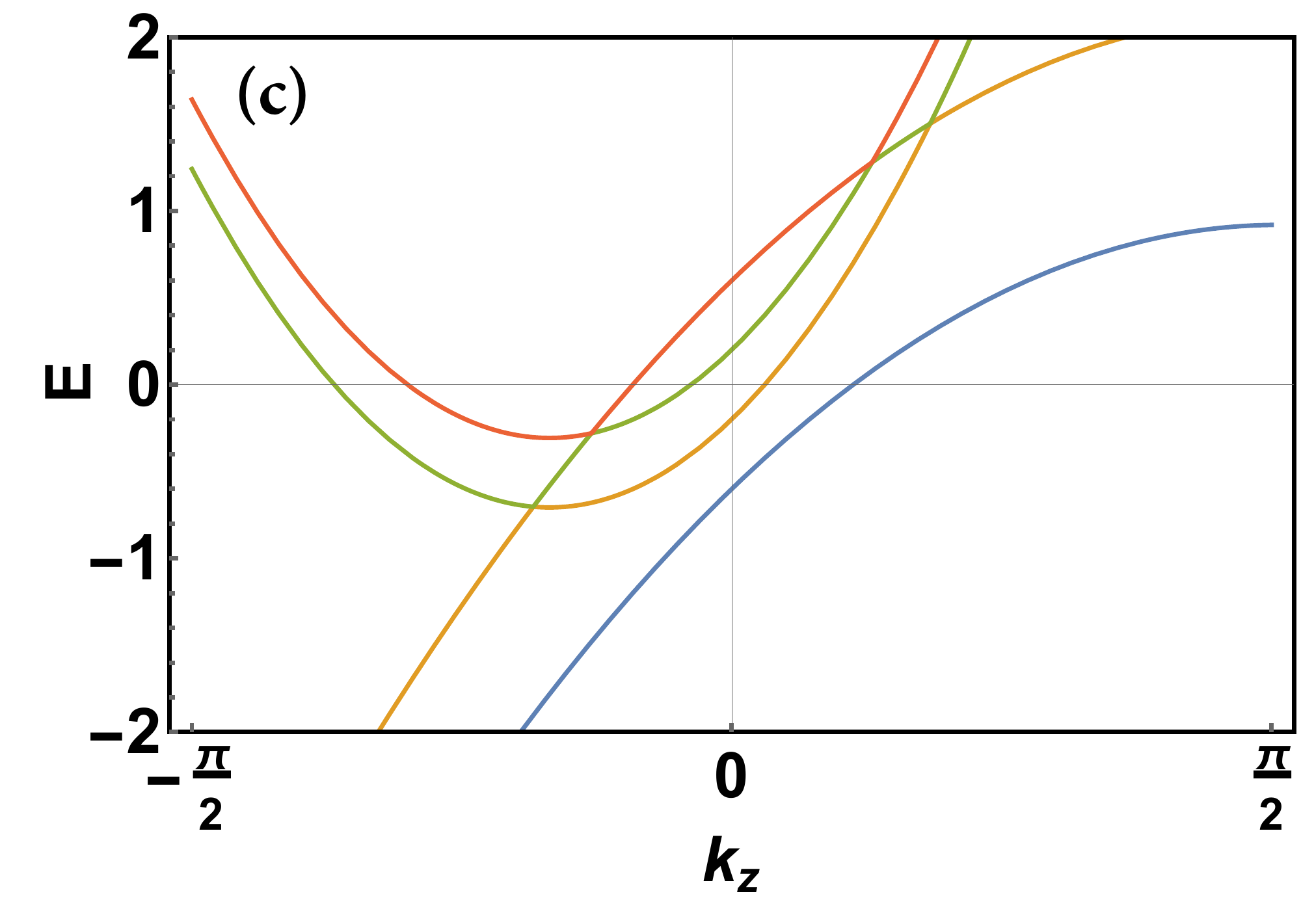}
\includegraphics[width=0.3\textwidth]{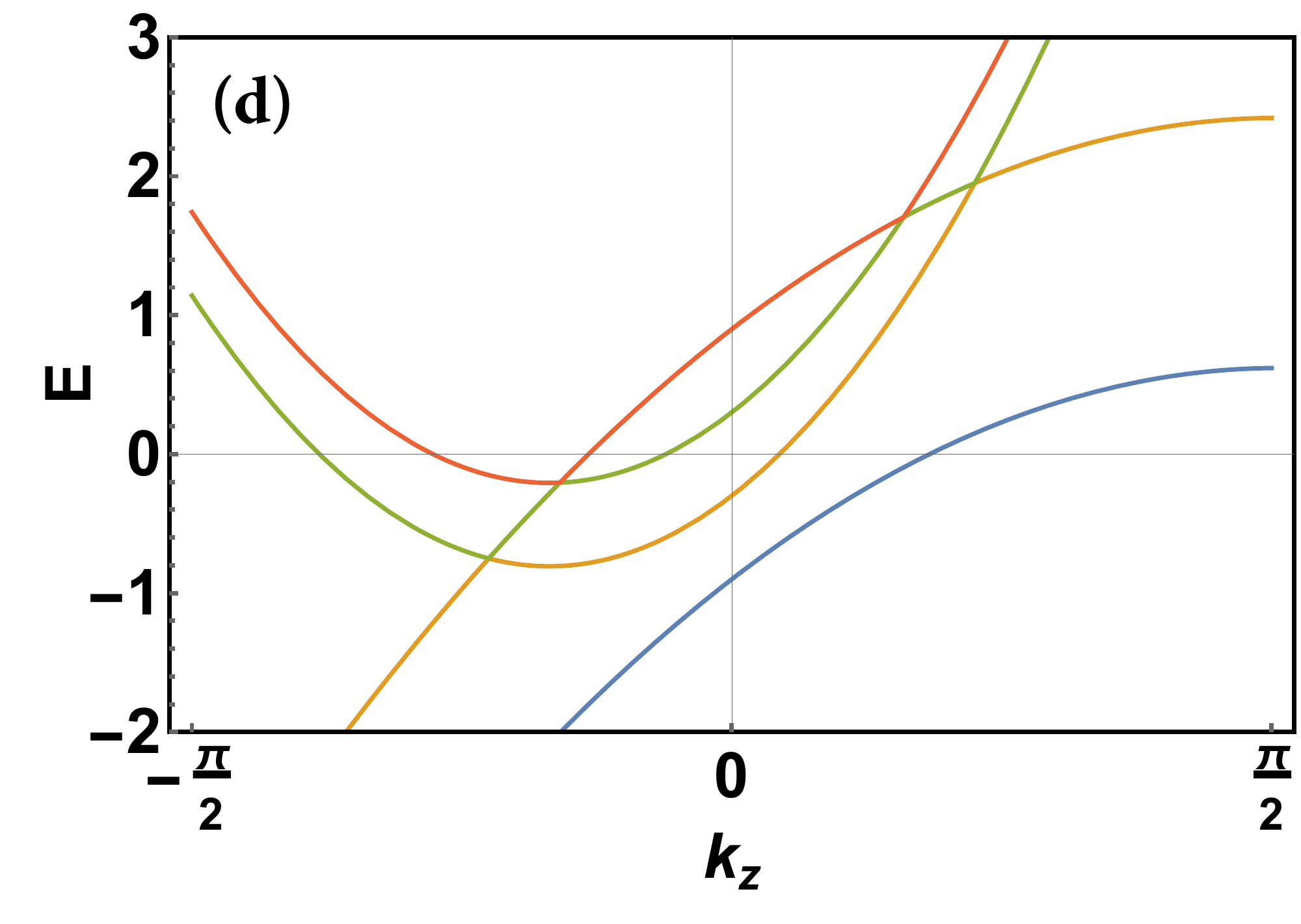}
\includegraphics[width=0.3\textwidth]{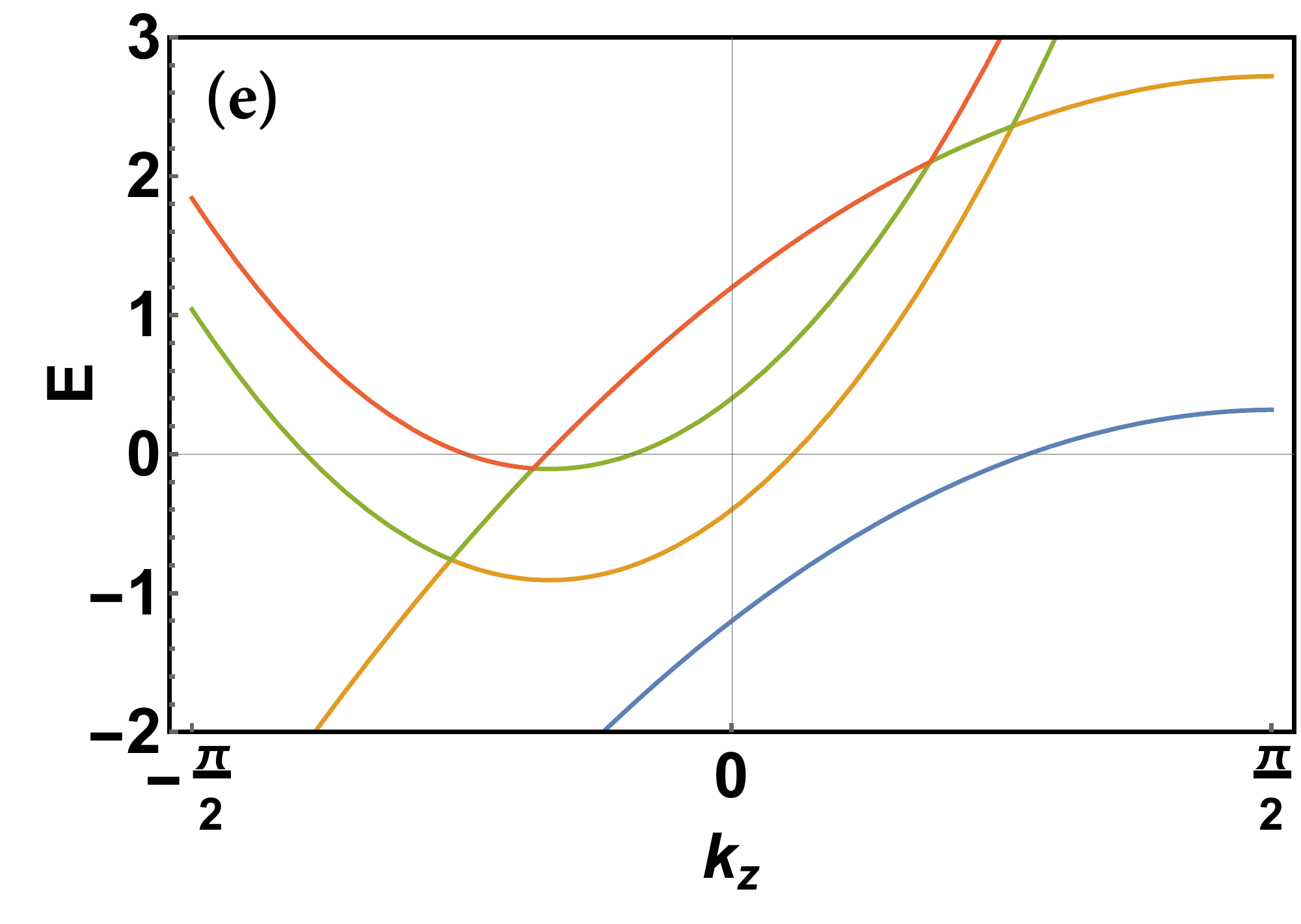}\includegraphics[width=0.3\textwidth]{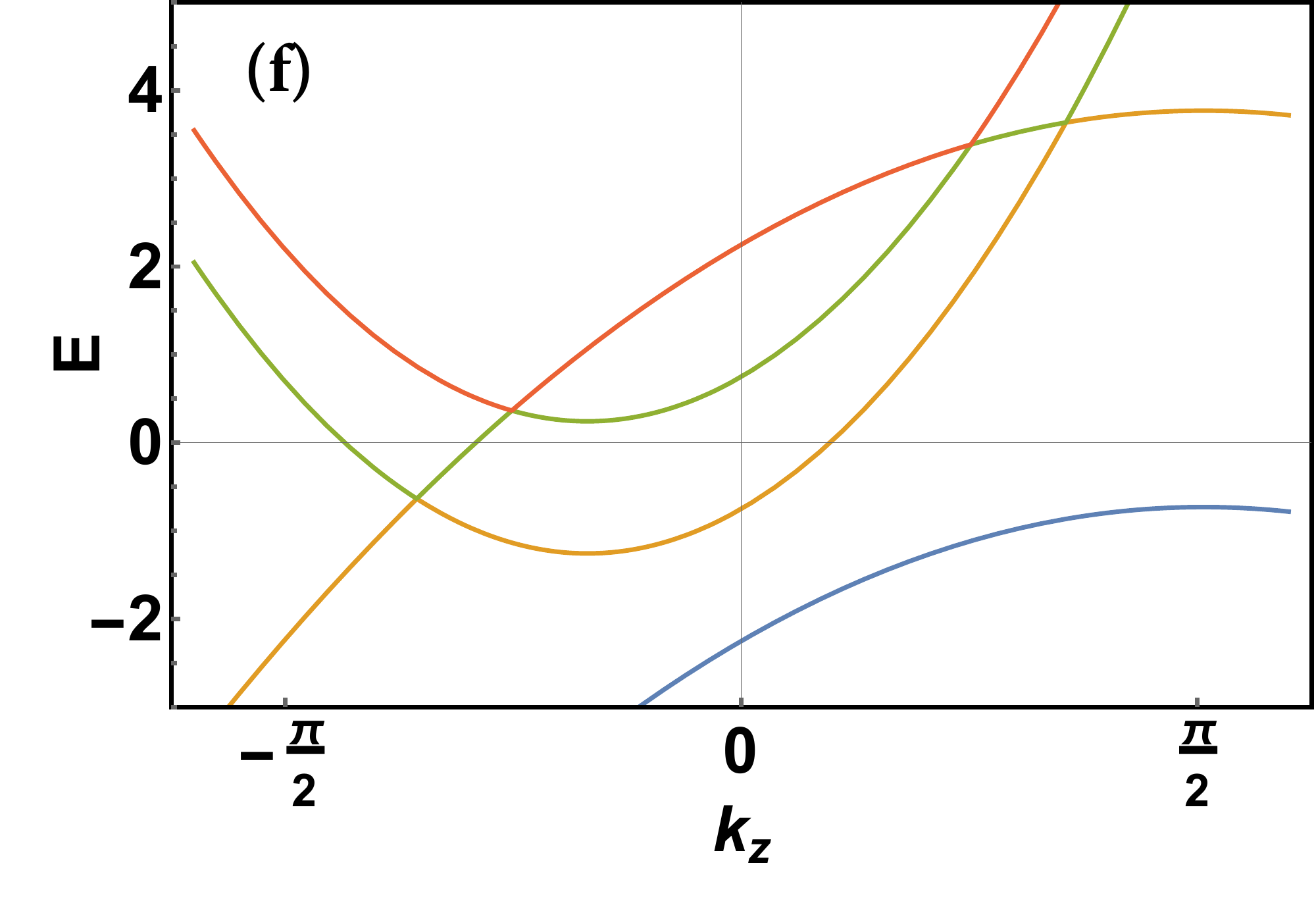}
\caption{The band structure of LSM with tilted QBT with kick strength of $\alpha=0.6$  and (a) $h=0$, (b) $h=0.2$, (c) $h=0.4$, (d) $h=0.6$, (e) $h=0.8$ and (f) $h=1.5$. $\lambda_1=\lambda_2=0.6$ and $\omega=20$ are used in all plots.}
\label{fig:kzG0}
\end{figure*}
\subsection{$\Lambda=k_z\Gamma_0$: Tilted QBT}

The second class of nonuniform kicks which we investigate in this work, is $\Lambda=k_z\Gamma_0$. The effective spectrum for such driving can be obtained as,
\begin{align}\label{tQBT}
    E_{\pm}(\vec{k})=(\lambda_{1}\mp2\lambda_{2})k^{2}+\frac{\alpha k_z}{T}.
\end{align}
The mere effect of such driving is to effectively tilt the QBT in $k_z$ direction as shown in Figure.~\ref{fig:kzG0}(a). However, the tilted QBT behaves very differently in the presence of an external perturbation. To clarify further, we rewrite the Eq.~(\ref{tQBT}) at $k_x=k_y=0$ as,
\begin{align}
     E_{\pm}(\vec{k})=&\,(\lambda_{1}\mp2\lambda_{2})k_z^{2}+\frac{\alpha k_z}{T}\cr
     =&\,(\lambda_{1}\mp2\lambda_{2})(k_z+\mathcal{A}_z)^{2}\cr
     =&\,(\lambda_{1}\mp2\lambda_{2})k_z^2+2(\lambda_{1}\mp2\lambda_{2})\mathcal{A}_z k_z+(\lambda_{1}\mp2\lambda_{2})\mathcal{A}^2_z\cr
     \simeq&\,(\lambda_{1}\mp2\lambda_{2})k_z^2+2(\lambda_{1}\mp2\lambda_{2})\mathcal{A}_z k_z,
\end{align}
where in the last line we assumed $\mathcal{A}_z<1$. Now by comparing the first and the last line of the above equation, we obtain $\mathcal{A}_z=\frac{\alpha}{2(\lambda_{1}\mp2\lambda_{2})T}$. Therefore, we can think of a tilted QBT as a QBT with a pseudo-electromagnetic potential $\mathcal{A}= (0, 0, \mathcal{A}_z)$ which is proportional to kick strength. In particular, interesting nontrivial topological phases can emerge by applying an external magnetic field in the presence of tilt, which otherwise are absent in a conventional QBT. As we mentioned previously, it is known that in the presence of a magnetic field, the QBT splits to multiple Weyl points. However, when the QBT is tilted, due to the competition between the external magnetic field and pseudo-magnetic field due to the $\mathcal{A}$ (proportional to kick strength) there are various regimes which different types of WSM phases can be generated. Starting with the weak field limit ($\alpha >> h$), there are two pairs of nodes, two double and two single nodes. In this regime, all Weyl nodes have a type-II nature (Figure.~\ref{fig:kzG0}(b)). By further increasing of the field, for intermediate fields ($h \lesssim \alpha$), hybrid Weyl pairs are realized where one of nodes in each of single and double pairs are type-II and the others are type-I. Interestingly, the realized hybrid WSM can survive up to a very strong magnetic field. This could be experimentally beneficent as it demonstrates that hybrid WSMs generated here, are accessible in a broad range of external fields. Similar to the $k_z\Gamma_5$ kick, we can get the analytical expression of energies and nodes position in the particle-hole limit,
\begin{align}
    E^{(1)}_{\pm}(\vec{k})=&\pm\frac{h}{2}+(2\lambda_2 k^2_z+\frac{\alpha k_z}{T}),\cr E^{(2)}_{\pm}(\vec{k})=&\pm\frac{3h}{2}-(2\lambda_2 k^2_z-\frac{\alpha k_z}{T}),
\end{align}
and,
\begin{align}
    k^{\pm}_{z,I}=\pm\sqrt{\frac{h}{2\lambda_2}},\,\,k^{\pm}_{z,II}=\pm\frac{\sqrt{h}}{2\sqrt{\lambda_2}}=\frac{k^{\pm}_{z,I}}{\sqrt{2}}.
\end{align}
Unlike the $k_z\Gamma_5$ kick, the presence of particle-hole symmetry does not prevent the generation of hybrid Weyl phases, because does not generate any other Dirac nodes and as we mentioned before only tilts the QBT. Moreover, the position of Weyl nodes are independent of kick strength and can be tuned solely using magnetic field.
\begin{figure}[htb]
    \centering
    \includegraphics[width=0.4\textwidth]{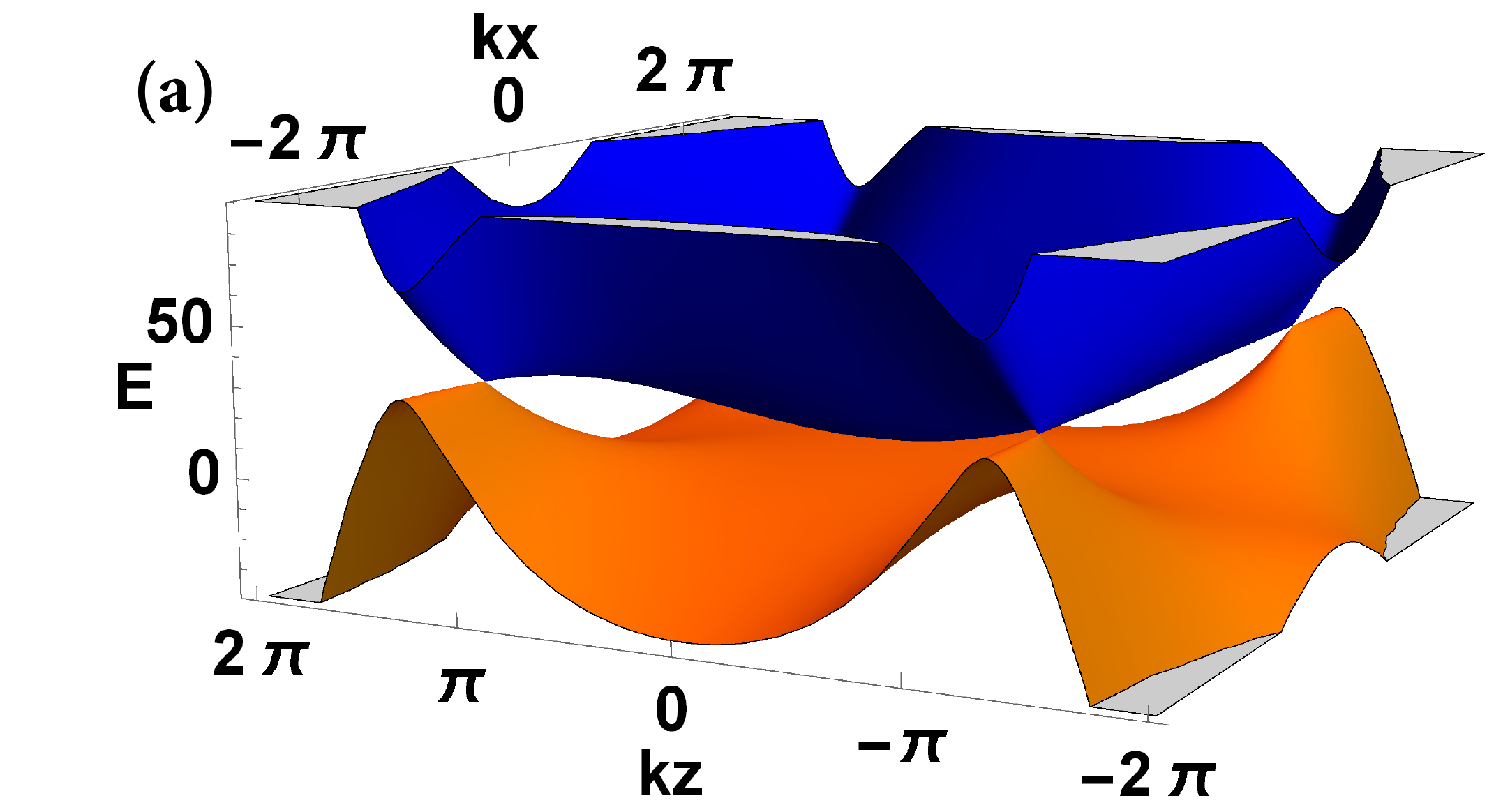}
    \includegraphics[width=0.4\textwidth]{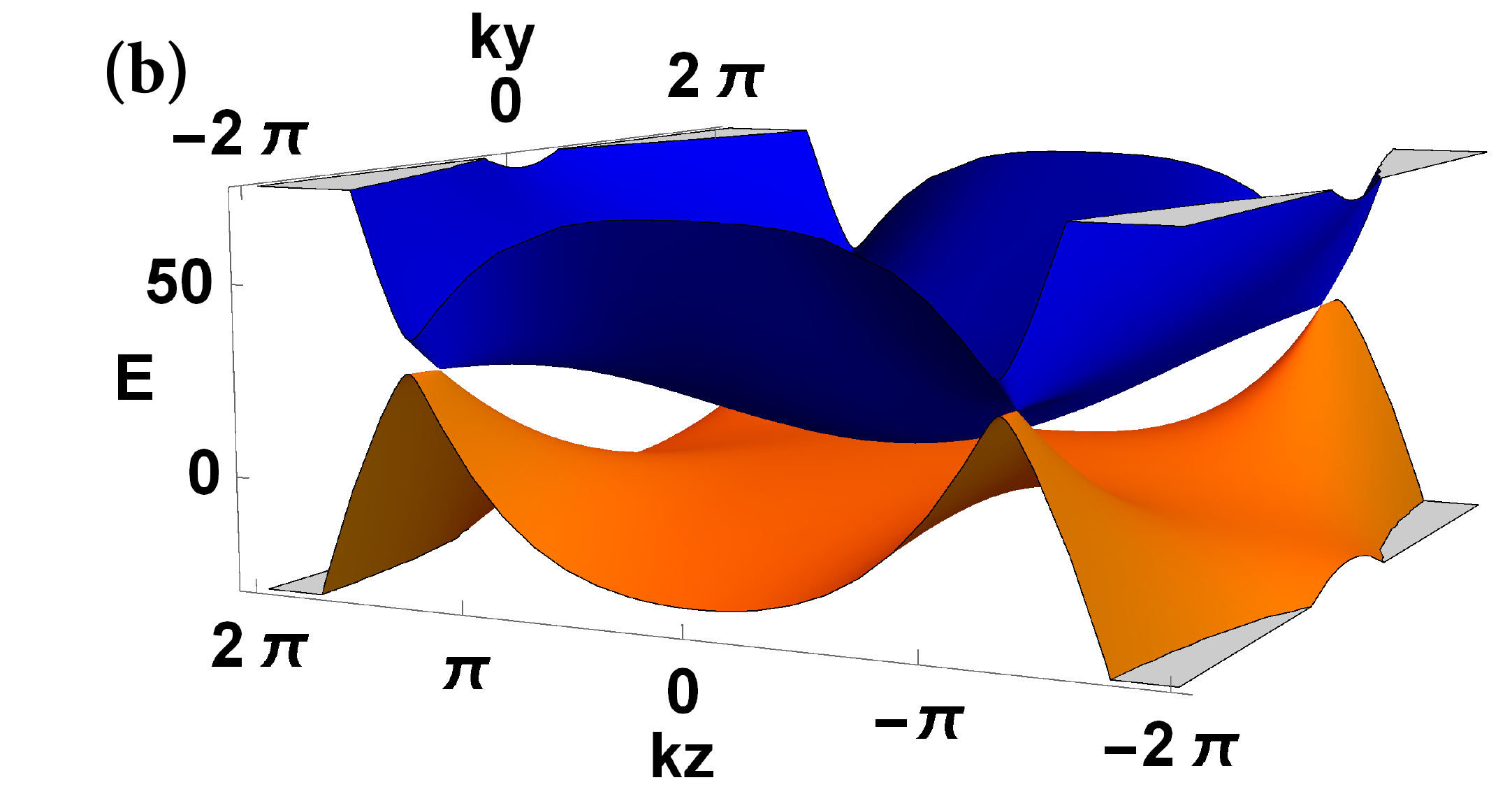}
    \caption{The bandstructure of LSM at $k_x-k_z$ plane ($k_y=0$) in presence of $k_z\Gamma_5$ smooth driving with $\alpha=5$. $\lambda_1=\lambda_2=0.6$  and $\omega=20$ are used.}
    \label{fig:kzG5smooth}
\end{figure}

\subsection{Smooth driving}
Here, we briefly discuss the case of smooth driving using Eq.~(\ref{effsmooth}) and make comparison with the results obtained by periodic kicking in previous sections. First of all, any type of perturbation which is proportional to the identity will not modify the system in the case of smooth driving, as it obviously commutes with Hamiltonian. However, the $\Lambda(\vec{k})=\alpha k_z\Gamma_5$ can modify the system when is applied via smooth driving. We obtain the effective Hamiltonian as,
\begin{align}
    H^{cos}_{eff}(\vec{k})=&H_0+\frac{\lambda_2\alpha^2k^2_z}{\omega}\bigg(2k_xk_z\{J_x,J_z\}+2k_yk_z\{J_y,J_z\}\cr
    +&(k^2_x-k^2_y)(J^2_x-J^2_y)+2k_xk_y\{J_x,J_y\}\bigg).
\end{align}
with energies,
\begin{align}
    E^{\pm}(\vec{k})=\lambda_1 k^2\pm \frac{\lambda_2\sqrt{f(\vec{k})\left[k^2_z\alpha^2-2\omega^2\right]+4k^4\omega^4}}{\omega^2},
\end{align}
where each $f(\vec{k})=3 (k_x^2+k_y^2)k_z^2 (k_x^2+k_y^2 +4k_z^2)\alpha^2$ and $E^{\pm}$ are doubly degenerate. Figure.~\ref{fig:kzG5smooth} shows the spectrum for $k_y=0$ plane, where four linear Dirac nodes coexist with a QBT at which is located at $\Gamma$ point. A similar plot can be obtained for $k_x=0$ plane, then the system also possesses line-nodes in $k_x-k_y$ plane at $k_z\neq 0$. Therefore, in addition to the coexistence of Dirac nodes and QBT at $k_x-k_z$ and $k_y-k_z$ planes, a smooth driving can lead to a richer phase diagram. This analysis was only for the purpose of comparison with kicked driving results, so a detailed analysis of such models will be left for future works.

\subsection{Effect of Lattice Regularization}

\begin{figure}[h]
    \centering
    \includegraphics[width=0.3\textwidth]{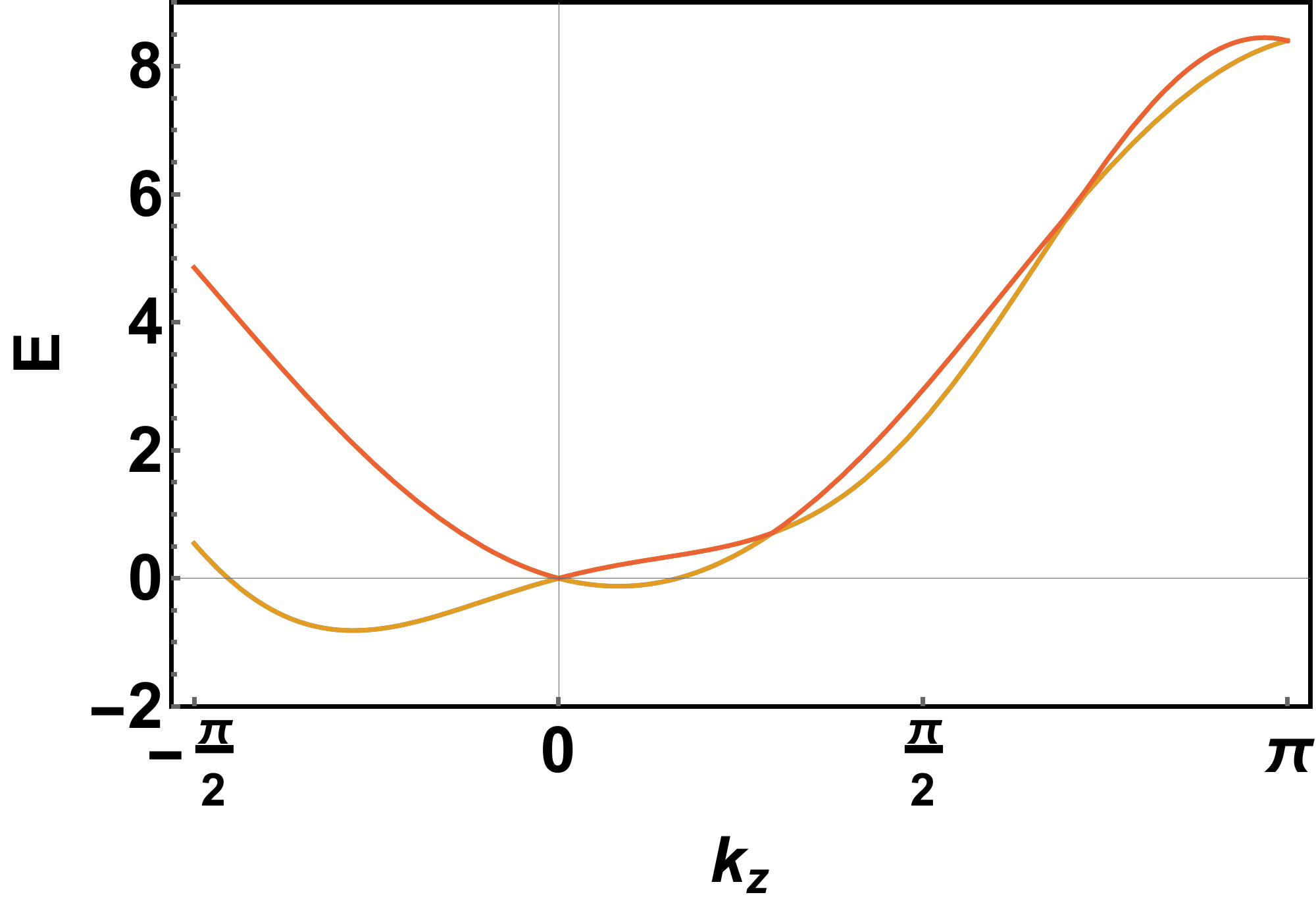}
    \caption{The $k_x=k_y=0$ cut of the bandstructure of LSM on a cubic lattice in presence of $\alpha\sin(k_z)\Gamma_5$ ($\alpha=0.3$) along the $k_z$ axis. $\lambda_1=\lambda_2=0.6$  and $\omega=20$ are used.}
    \label{fig:latt}
\end{figure}

Lets now look at the effect of lattice regularization. One of the typical effect of lattice versus continuum model is the appearance of more nodes, usually at the boundaries of the Brillioun Zone \cite{Ghorashi2018}. However, we show that the main features discussed in the previous subsections survives in lattice models. For the sake of brevity we restrict ourselves to $h=0$ limit. We consider a cubic lattice, which captures the physics of continuum model for $\alpha,\omega,\vec{k}<<1$. In the $k_z$ cut (Figure.~\ref{fig:latt}), we find: (i) two nodes at BZ boundaries, $k_z=\pm \pi$, as we expected, (ii) one at the $\Gamma$ point with quadratic dispersion in $k_i \perp k_z$, in agreement with the result of continuum model, (iii) however, due to lattice regularization there can be two nodes (instead of the node at $k_z=\frac{\alpha}{2\lambda_2 T}$ in the continuum model) with linear dispersion away from the boundaries and the $\Gamma$ point, which are located at $k_z=\sin^{-1}(\frac{\alpha}{2\lambda_2 T}), k_z=-\sin^{-1}(\frac{\alpha}{2\lambda_2 T})+\pi$. Figure.~\ref{fig:latt}, shows the lattice bandstructure along $k_z$ for the case of $k_z\Gamma_5$ driving. Therefore, we confirm that the generation of the hybrid dispersion Dirac semimetal (and consequently Weyl semimetal) persists in the lattice model.

\section{DISCUSSION AND CONCLUDING REMARKS}

We have proposed a dynamical way to realize hybrid Dirac and Weyl semimetals in a three-dimensional Luttinger semimetal via applying a nonuniform (momentum-dependent) periodic $\delta$-kick. We explicitly demonstrated this through two examples of nonuniform kicking which break both inversion and time-reversal symmetries while preserving their combinations. We have identified the first example of an unusual hybrid Dirac semimetal phase where two nodes not only have different types but also have different dispersions ( a linear Dirac and QBT coexist). Then by applying an external magnetic field we demonstrated the emergence of hybrid Weyl semimetals. Next, we found that the combination of a tilted QBT with an external magnetic field provides a promising setup for generation of hybrid Weyl semimetals. Moreover, by interpreting the tilted QBT as a QBT with emergent psudomagnetic field proportional to kick strength, we discussed the interplay between the kick strength and the external magnetic field. \\
\indent We note that the experimental realization of models discussed in this work can be difficult, specifically for the case of uniform strain which requires very fast time scales for periodic driving. Despite all the difficulties there are some proposals for the fast dynamical generation of strain \cite{straindyn1,straindyn2,kickgraphene}. However, the possibility of interpreting the $k_z\Gamma_5$ kick as an applied spin current (or proportional to it) along the $z$-direction is potentially a promising route, considering the recent developments in the field of ultrafast spintronics \cite{ultrafastspin}. Moreover, the 2D QBT has been already proposed to be realized in optical lattices \cite{QBToptical}, therefore, in practice a 3D QBT could be realized in optical lattices as a potential experimental setup where parameters can be tuned at will.\\
\indent The LSMs can describe the low-energy physics of many experimental candidates. Therefore, this work opens up a promising way for the realization of various hybrid Dirac and Weyl semimetals. Therefore, it could motivate further studies on the lacking investigation of physical properties of the hybrid Weyl semimetals. \\
Some of the possible future directions are: the investigation of bulk and surface transport properties in the hybrid dispersion Dirac semimetals introduced here, in particular, how some of the most interesting properties of a typical Dirac semimetal, such as transport anomalies \cite{review1}, would be different. Moreover, it would be interesting to see whether other multi-band systems with higher-spin can show similar physics.

\section*{Acknowledgement}
We thank Matthew Foster and Bitan Roy for useful comments. We also acknowledge useful comments and suggestions from the anonymous referees which helped to improve this manuscript. This work was supported by the U.S. Army Research Office Grant No.W911NF-18-1-0290. we also acknowledge partial support from NSF CAREER Grant No. DMR-1455233 and ONR Grant No. ONR-N00014-16-1-3158.

\bibliography{kickQBT}




\end{document}